\documentclass[12pt,a4paper]{article}
\pdfoutput=1  
\usepackage{amsmath,amsfonts,amssymb,amsbsy}
\usepackage{graphicx}
\usepackage{color}
\usepackage{booktabs}
\usepackage{multirow}
\usepackage{multicol}
\usepackage{subfigure}
\usepackage{alpha}
\usepackage{macros_alpha}
\providecommand*{\psibar}{\ensuremath{\overline{\psi}}}

\renewcommand{\strange}{\mathrm{strange}}

\begin{document}

\preprintno{%
\vspace{0.9cm}
CP3-Origins-2018-001 DNRF90\\
IFT-UAM/CSIC-18-010\\
FTUAM-18-3
}

\title{%
Tuning the Hybrid Monte Carlo algorithm using 
molecular dynamics forces' variances\\
\vspace{.5cm}
}

\author{A. Bussone$^{a,b,c}$, M. Della Morte$^a$, V. Drach$^{d}$, 
and C. Pica$\,^a$ \\
\vspace{1.5cm}
\begin{small}
{$^a$}{CP$^3$-Origins, University of Southern Denmark, Campusvej 55, 5230 Odense, Denmark\\}
{$^b$}{Department of Theoretical Physics, Universidad Aut\'onoma de Madrid, E-28049 Madrid, Spain\\}
{$^c$}{Instituto de F\'{\i}sica Te\'orica UAM-CSIC, c/ Nicol\'as Cabrera 13-15, Universidad Aut\'onoma de Madrid, E-28049 Madrid, Spain\\}
{$^d$}{Centre for Mathematical Sciences, Plymouth University,
Plymouth, PL4 8AA, UK \\}
\vspace{-3.7cm}
\end{small}
}
\begin{abstract}
Within the HMC algorithm,
we discuss how, by using the shadow Hamiltonian and the Poisson brackets, one can achieve
a simple factorization in the dependence of the Hamiltonian violations upon either the algorithmic parameters
or the parameters specifying the integrator.
We consider the simplest case of a second order (nested) Omelyan \textcolor{black}{integrator} and one level of 
Hasenbusch splitting of the determinant for the simulations of a  QCD-like theory (with gauge group SU(2)).
Given the specific choice of the integrator, the Poisson brackets reduce to the variances
of the molecular dynamics forces. We show how the factorization can be used to optimize in a very 
economical and simple way both the algorithmic and the \textcolor{black}{integrator} parameters with good accuracy.
\end{abstract}


\maketitle 

\section{Introduction}
Gauge theories formulated on Euclidean lattices can be treated as statistical systems and are amenable
to numerical simulations. When matter fields (scalars or fermions) are also present, the gauge-field configurations
are typically generated using molecular dynamics algorithms, which come with different variations of the
Hybrid Monte Carlo (HMC) algorithm~\cite{Duane:1987de}.
Such algorithms are characterized by  a large number of parameters, whose optimal choices depend on the model and the regime
(e.g., concerning masses and volumes) considered.

The case of QCD has been extensively studied because of its obvious phenomenological relevance.
Roughly speaking HMC algorithms can be classified according to the factorization of the quark determinant
adopted (typically either mass-preconditioning~\cite{Hasenbusch:2001ne}, or 
domain-decomposition~\cite{Luscher:2005rx}, or rational factorization~\cite{Clark:2006fx})
and the symplectic integrator(s) used, the simplest being the leapfrog integrator, and one of the most
popular being the second order minimum norm integrator or Omelyan-integrator~\cite{Omelyanint,Takaishi:2005tz}.
The two choices are actually connected. The factorization of the determinant translates into a splitting
of the fermionic forces in the molecular dynamics, and depending on the hierarchy of such forces various
nested integrators can be used where each force is integrated along a trajectory with a different 
time-step~\cite{Luscher:2005rx,Urbach:2005ji}.
It is well known that for symplectic integrators a shadow Hamiltonian exists, which is conserved
along the trajectory. This observation can be exploited to construct efficient integrators, as 
done in~\cite{Clark:2008gh,Clark:2011ir}. In short, the idea is to minimize the fluctuations in the difference
between the Hamiltonian of the HMC and the shadow Hamiltonian along the trajectory.
Since the latter is constant, the procedure clearly minimizes, in particular, the difference between the initial and
the final Hamiltonian, hence increasing the acceptance.

The relation between the shadow Hamiltonian and the Hamiltonian of the system involves complicated functions 
(Poisson brackets) of the HMC-momenta and of the field variables, however for a particular choice of the integrator, 
these expressions simplify and one is left with basically the variance of the fermionic forces.
We specialize exactly to that choice, that we detail in the following, since in this way the forces computed
(and stored in some form) from previous simulations can be used to find the optimal choice of parameters
for the factorization of the determinant and for the step-sizes in the hierarchical integration scheme.

This is particularly relevant for the case of QCD-like theories which may 
provide viable strongly-interacting extensions of the Standard Model and 
therefore have to be studied within a non-perturbative approach. 
The field is now moving towards accurate quantitative predictions, and
that requires efficient algorithms. Many such models for example
feature matter fields in representations of the gauge group 
(not necessarily SU(3)) which are not the fundamental one.
In this case the hierarchy between the gauge and fermionic forces may be 
very different from the case of QCD and that implies that the 
algorithmic optimization has to be repeated.

As mentioned, the approach we discuss here is rather economical, and although
restricted to a particular integrator, it can be easily applied
to different QCD-like models and to different factorizations of the determinant.
We test the method here by considering an SU(2) gauge theory with one doublet
of fermions in the fundamental representation and by adopting the (single)
Hasenbusch factorization of the determinant\cite{Hasenbusch:2001ne}.
Within this framework we show that we can predict the dependence (e.g.) of
the acceptance on the different parameters at the 10\% level. 
A preliminary account of this study has appeared in~\cite{Bussone:2016pmq}.

The paper is organized as follows: in the 
next Section  we recall some basic properties of the HMC algorithm, of integrators and of the shadow Hamiltonians, 
in Section 3 we discuss Creutz's formula for the acceptance and define
our optimization procedure, Section 4 contains tests and results, and in Section 5 we present
our conclusions and outlook.

\section{(Not so) Basic facts about symplectic integrators and HMC}

We briefly introduce the concepts about integrators and the Hybrid Monte Carlo
algorithm that are necessary in order to present our optimization strategy.
The reader familiar with dynamical simulations of lattice QCD can probably
skip this Section.

\subsection{The analytical mechanics case}
The example of the numerical integration of the equations of motion for a classical Hamiltonian system
is quite instructive also in view of the application to gauge theories.
\textcolor{black}{For the latter however one needs to formulate Hamiltonian mechanics on Lie groups, as
explained for example in~\cite{Kennedy:2012gk}.}
The classical system is described by an Hamiltonian $H$ which,
by adopting standard notation, we write as $H(q,p)=\frac{1}{2} p^2 + S(q)$
and further define $T(p)$ as $\frac{1}{2} p^2$. The term $S(q)$, which only depends
on the coordinates $q$ (to be later identified with the gauge field) may sometimes
be referred to as the potential. A symplectic (i.e. phase-space volume preserving)
integrator is constructed by discretizing the time evolution operator
\begin{equation}
\exp\left( \tau \frac{d}{dt}\right)=\exp{\left(\tau\left( \frac{dp}{dt} \frac{\partial}{\partial p}
+ \frac{dq}{dt} \frac{\partial}{\partial q}  \right)\right)} \equiv \exp{\left( \tau \hat{H}\right)} \;,
\end{equation}
with $\hat{H}$ given by
\begin{equation}
 \hat{H}=-\frac{\partial H}{\partial q}\frac{\partial}{\partial p}     +  
\frac{\partial H}{\partial p}\frac{\partial}{\partial q} \equiv -S_q \frac{\partial}{\partial p} 
+T_p \frac{\partial}{\partial q} \equiv \hat{S} +  \hat{T} \;.
\end{equation}
It is easy to see, given that $T$ only depends on $p$ and $S$ on $q$, that the action of 
the time evolution operator on a function of coordinates and momenta simply implements Hamilton
equations.

Upon discretizing time in steps of size $\delta \tau$ in view of computational applications, 
one is forced to introduce numerical integrators. For HMC those additionally need to be
symmetric in time in order to satisfy detailed balance.
A common choice\footnote{That is actually the only choice we consider here.} is the 
second order minimum norm (or Omelyan) integrator~\cite{Omelyanint},
for which the evolution over a trajectory of length $\tau$ can be written as
\begin{equation}
U_{\rm Om}(\delta\tau,\alpha)^{\tau/\delta\tau}=\left(e^{\alpha \delta\tau \hat{S}} e^{\frac{\delta\tau}{2} \hat{T}}
e^{\delta\tau (1-2\alpha)\hat{S}}  e^{\frac{\delta\tau}{2} \hat{T}} e^{\alpha \delta\tau \hat{S}} 
   \right)^{\tau/\delta\tau}\;,
\label{eq:Omelyan2onescale}
\end{equation}
\textcolor{black}{where $\alpha$ is a free parameter, typically chosen between 0 and 1/2 for forward integrators.}
By applying the Baker-Campbell-Hausdorff formula,
the operator above can be written as $\exp\left(\tau\hat{\tilde{H}}\right)$, with
\begin{equation}
\hat{\tilde{H}}=\hat{H}+   \left(\frac{6\alpha^2-6\alpha+1}{12}\left[\hat{S},\left[\hat{S},\hat{T}\right]\right]   +\frac{1-6\alpha}{24}  
\left[\hat{T},\left[\hat{S},\hat{T}\right]\right]\right)\delta\tau^2
 + O\left(\delta\tau^4\right) \;.
\end{equation}
As discussed in~\cite{Clark:2007ffa} to each symplectic integrator corresponds
an exactly conserved shadow Hamiltonian $\tilde{H}$, which, to a given order
in $\delta\tau$, can be obtained by replacing the commutators $\left[\hat{A},\hat{B}\right]$ in the expansion
of $\hat{\tilde{H}}$
with the Poisson brackets  
\[ \{A,B\}=\frac{\partial A}{\partial q} \frac{\partial B}{\partial p} - 
\frac{\partial A}{\partial p} \frac{\partial B}{\partial q} \;.\]
In the case of the Omelyan integrator considered here that yields
\begin{equation}
{\tilde{H}}={H}+   \left(\frac{6\alpha^2-6\alpha+1}{12}\left\{{S},\left\{{S},{T}\right\}\right\}   +\frac{1-6\alpha}{24}  
\left\{{T},\left\{{S},{T}\right\}\right\}\right)\delta\tau^2
 + O\left(\delta\tau^4\right) \;.
\label{eq:mech1}
\end{equation}
It is worth having a closer look at such Poisson brackets. One easily obtains
\begin{eqnarray}
\label{eq:mech2}
\left\{{S},\left\{{S},{T}\right\}\right\} & = & S_q^2=F^2 \;,  \\
\left\{{T},\left\{{S},{T}\right\}\right\} & = & -p^2 S_{qq} \;,
\label{eq:mech3}
\end{eqnarray}
where we have introduced the driving force $F=-S_q$ that also appears in Hamilton's equation
$dp/dt=F$. The other Poisson bracket involves instead the second derivative of the potential,
and may be computationally expensive, especially in the case of QCD-like theories 
(see~\cite{Kennedy:2012gk} for a discussion and a numerical approach).
Clearly the choice $\alpha=1/6$ simplifies the expression of
the shadow Hamiltonian up to $O\left(\delta\tau^4\right)$.
In particular, the difference $\delta H = \tilde{H}-H$, in that case, only depends
on the driving forces already computed in order to
integrate the equations of motion.

\subsection{Mass-preconditioned HMC and multi time-scale integrators}{\label{onHMC}}
We now consider the case of lattice gauge theories with fermionic matter
(QCD-like theories) and we briefly introduce the mass-preconditioned 
algorithm~\cite{Hasenbusch:2001ne}  that we used in this study. 
Let us consider the massive Dirac operator in the Wilson regularization
\begin{equation}
D_W=D+m_0\;, \quad {\rm and} \;\; Q=\gamma_5 D_W=Q^\dagger\;.
\end{equation}
The version of Hasenbusch's or mass preconditiong adopted here amounts to introducing an
infrared parameter $\mu$ (a mass parameter) and then decomposing
the Dirac operator as
\begin{equation}
D_W=\left(D_W+\mu \right) \times \left( D_W+\mu\right)^{-1}D_W\equiv D_2(\mu) \times D_1(\mu)\;,
\label{eq:musplit}
\end{equation}
and similarly for $Q$.
The parameter $\mu$ can be chosen such that the two operators
have, on average, the same condition number $\sqrt{\kappa}$, where $\kappa$
is the average condition number of the initial matrix $Q$ (see~\cite{DellaMorte:2003jj} for the case of two degenerate flavors).
Given that the number of conjugate gradient iterations needed in order to invert
an Hermitean operator on a given source (as it is needed in the computation of  the driving
forces) is proportional to the condition number, the mass preconditioning may provide a
substantial speed-up of simulations.
A different, perhaps more efficient strategy, consists in tuning $\mu$ such
that the operator $\left(D_W+\mu\right)$ is rather cheap to invert but gives the largest
contribution to the driving force of the momenta, whereas the other factor 
remains more expensive and contributes little to the force.
For the case of lattice QCD the existence of such a choice has
been first discussed, on the basis of numerical data, in~\cite{Urbach:2005ji}.
As proposed there, it is then rather natural to introduce a multilevel
integration scheme, where the different forces are integrated using
different time-steps $\delta \tau$ inversely proportional to the
magnitude of the force. We are going to discuss in some detail the origin and the properties
of the different force contributions in the remaining part of this section.
That will naturally lead to our optimization procedure, which is
based on the variance of the forces, rather than on their magnitude.

A couple of remarks are in order on the splitting we have chosen in eq.~\ref{eq:musplit}.
By looking at the cases $\mu=0$ and $\mu \to \infty$ one observes
that $D_1(\mu)$ and $D_2(\mu)$ exchange role in these limits and therefore
the hierarchy between the corresponding forces must flip around
some critical value of the $\mu$ parameter.\footnote{Notice that the molecular
dynamics evolution in the HMC algorithm is independent from the normalization
of the fermionic action (as it should be). Such a normalization is absorbed in the
generation of the pseudo-fermion fields in the heatbath at the
beginning of each trajectory~\cite{Duane:1987de}.} 
Secondly, notice that the operators $D_1(\mu)$ and $D_2(\mu)$ are not invariant under
the exchange $\mu \leftrightarrow -\mu$, also when the case of two generate flavors
is considered. 

We write the Hamiltonian in the HMC evolution for the case of two degenerate fermions as
\begin{eqnarray}
\label{eq:SGS1S2}
H &=& \frac{1}{2} \sum_{x,\mu} \Tr \left[\pi_\mu(x)^2\right] + S_{\rm G}(U)+ 
\phi_1^\dagger  \left(Q^{-1}D_2^\dagger D_2  Q^{-1} \right)\phi_1 +
\phi_2^\dagger\left[\left(\gamma_5 D_2\right)^2\right]^{-1} \phi_2
\;,  \nonumber \\
& \equiv & \frac{1}{2} (\pi,\pi) + S_{\rm G}(U)+ S_1(U,\phi_1^\dagger,\phi_1)+ S_2(U,\phi_2^\dagger,\phi_2) \;,
\end{eqnarray}
with the momenta $\pi$ defined as $\pi_\mu(x)=\pi_\mu^a(x) T^a_f$, with $T^a_f$ the hermitean
generators of the Lie algebra ${\mathfrak{su}}(2)$ in
 the fundamental representation of the gauge 
group.\footnote{We use the normalization $\Tr(T^a_f T^b_f)=\frac{1}{2}\delta_{ab}$.}
 $S_{\rm G}(U)$ is the Wilson plaquette action and the pseudo-fermion fields $\phi_{1,2}$ are introduced in order 
to re-express, upon integration, the fermionic determinant. They are generated through an heatbath step at the beginning of
each trajectory and kept fixed during the evolution.

The  update by a time-step $\delta \tau$ of the fields $\pi$ and $U$ (i.e., the action of the operators $e^{\delta \tau \hat{S}}$
and $e^{\delta \tau \hat{T}}$ respectively), can be defined as
\begin{eqnarray}
\pi_\mu(x)  & \rightarrow & \pi'_\mu(x)=\pi_\mu(x)+\delta\tau \sum_i F_i^\mu(x)  \;, \quad i={\rm G},1,2 \nonumber \\
U_\mu(x) & \rightarrow & U'_\mu(x)=e^{i\delta \tau \pi_\mu(x)}U_\mu(x) \;,
\end{eqnarray}
\textcolor{black}{where the forces 
$F_i^\mu(x) = F_i^{a\mu}(x) T^a_f$ 
are 
determined by loosely speaking deriving the action with respect to the gauge fields, more formally:}
\begin{equation}
F_i^{a\mu}(x) = - \lim_{\varepsilon \to 0}\left\{\frac{1}{\varepsilon} \left( S_i|_{U\to U_\varepsilon} -S_i\right)\right\}  \;,
\quad (U_\varepsilon)_\mu(x)=e^{i\varepsilon(x,\mu) T^a_f}U_\mu(x)\;,
\end{equation}
\textcolor{black}{
with real variables $\varepsilon(x,\mu)$. For more details see for example~\cite{Luscher:2010ae}.}

In a multilevel (or nested) integration scheme instead of a common time-step $\delta \tau$
one introduces a different time-scale $\delta \tau_i$ for each force contribution
and a corresponding operator $e^{\delta\tau_i \hat{S}_i}$. For example, the two-level second order
Omelyan integrator, assuming that there are only two contributions to the force and that
those are integrated using $\delta \tau_1=\delta \tau$ and $\delta \tau_2=\delta \tau/m$, is
obtained by the following replacements
\begin{equation}
\hat{S} \rightarrow \hat{S}_1\;, \quad
 e^{\frac{\delta \tau}{2}\hat{T}} \rightarrow  \left( e^{\alpha \frac{\delta\tau}{2m}\hat{S}_2}   e^{ \frac{\delta\tau}{4m}\hat{T}}
e^{\frac{\delta\tau}{2m}(1-2\alpha)\hat{S}_2}    e^{ \frac{\delta\tau}{4m}\hat{T}}  e^{\alpha \frac{\delta\tau}{2m}\hat{S}_2}   \right)^m\;,
\end{equation}
in eq.~\ref{eq:Omelyan2onescale}. The shadow Hamiltonian to $O(\delta \tau^4)$ then reads
\begin{eqnarray}
{\tilde{H}}={H} &+&   \frac{6\alpha^2-6\alpha+1}{12}  \left(\left\{{S_1},\left\{{S_1},{T}\right\}\right\}  +\frac{1}{4m^2}
\left\{{S_2},\left\{{S_2},{T}\right\}\right\} \right) \delta\tau^2 \\
&+&\frac{1-6\alpha}{24}  \left(
\left\{{T},\left\{{S_1},{T}\right\}\right\}+\frac{1}{4m^2}
\left\{{T},\left\{{S_2},{T}\right\}\right\}+
\left\{{S_1},\left\{{S_2},{T}\right\}\right\}
  \right)\delta\tau^2 \;. \nonumber
\end{eqnarray}
One can construct higher level second order Omelyan integrators in a similar way.
The generic expression for $\tilde{H}$ that we could obtain, can be cast in the form
\begin{eqnarray}
\tilde{H} = H &+&\tau^2 \sum_{i} \left[ \frac{6\alpha^2-6\alpha+1}{3} 
\left( \frac{\left\{{S_i},\left\{{S_i},{T}\right\}\right\}}{\prod_{k=1}^i (2n_k)^2}\right)  \right. \\
&+& \left. 
\frac{1-6\alpha}{6} 
\left(\frac{\left\{{T},\left\{{S_i},{T}\right\}\right\}
+ \left\{{S_i},\left\{\sum_{j>i}{S_j},{T}\right\}\right\}    }{\prod_{k=1}^i (2n_k)^2}   
 \right)  \right] + O(\delta \tau^4)\;, \nonumber
\end{eqnarray}
with $\delta \tau=\tau/n_1=\delta\tau_1$, where $\tau$ is the trajectory length. The step-sizes for the inner levels are given by 
$\delta \tau_i= \tau/(\prod_{k=1}^in_k)$.

We used a three-level integration scheme with the force $F_1$ on the coarsest scale $\delta \tau=\tau/n$,
$F_2$ on $\delta\tau_2=\delta \tau/m$ and the gauge force $F_{\rm G}$  on the innermost level with time-step 
$\delta \tau_2/k$. The corresponding shadow Hamiltonian, setting $\alpha=1/6$ in each level, is given by
\begin{equation}
\tilde{H}=H+\frac{\delta \tau^2}{72}\sum_{x,\mu,a} \left[  T_{R,1} (F_1^{a\mu}(x))^2 + \frac{T_{R,2} (F_2^{a\mu}(x))^2 }{4m^2} +
\frac{T_{R,\rm G} (F_{\rm G}^{a\mu}(x))^2 }{16m^2k^2}
 \right] + O(\delta \tau^4) \;,
\end{equation}
where the $T_{R,i}$ are the squared norms of the generators in the relevant representation for the conjugate momenta $\pi$.
We stick to the fundamental representation in this study. In order to make contact with the Hamiltonian mechanics case we rewrite
the above equation in the form
\begin{equation}
\tilde{H}\equiv H+\frac{\delta \tau^2}{72} \left( \left|{\mathcal{F}}_1\right|^2 + 
\frac{ \left|{\mathcal{F}}_2\right|^2 }{4m^2}  + \frac{ \left|{\mathcal{F}}_{\rm G}\right|^2 }{16m^2k^2}      \right)
+ O(\delta \tau^4) \;,
\label{eq:calF}
\end{equation}
which indeed generalizes eqs.~\ref{eq:mech1} and~\ref{eq:mech2} (considering $\alpha=1/6$). 

At this point we need to relate $\Delta H$ (the change in the Hamiltonian over a trajectory) 
to $\Delta(\delta H)$, with $\delta H = \tilde{H}-H$. We are actually mostly interested
in the relation between the corresponding variances, as Var$(\Delta H)$ controls the
HMC acceptance. We recall these relations in the next Section.

\section{Cost definition and parameters optimization}
There are clearly many possible definitions of the cost of an HMC simulation, on general grounds though
they should all at least be inversely proportional to the 
acceptance and directly proportional to the computational cost.

The acceptance probability $P_{\rm acc}$  is related to the variance of $\Delta H$
through the Creutz formula~\cite{Creutz:1988wv,Gupta:1990ka}
\begin{equation}
P_{\rm acc}(\Delta H)= {\rm erfc}\left( \sqrt{{\rm Var}(\Delta H)/8} \right)
\;,
\label{eq:pacc}
\end{equation}
which is derived from the reversibility relation $\langle e^{-\Delta H} \rangle=1$
(where $\langle\dots\rangle$ indicates the average over the gauge configurations)
via a cumulant expansion, and it is therefore valid for $P_{\rm acc}$ close to 1.

From the fact that the shadow Hamiltonian is conserved along a trajectory,
it immediately follows that  Var$(\Delta H)=$~Var$[\Delta(\delta H)]$.
By further assuming that for long enough trajectories the initial and final values
of $\delta H$ are independently extracted from the same distribution  $\propto e^{-H}$, one finally obtains
\begin{equation}
{\rm Var}(\Delta H)={\rm Var}[\Delta(\delta H)]= 2 {\rm Var}(\delta H)\;,
\label{eq:varianceequalities}
\end{equation}
as discussed in~\cite{Clark:2010qw}.

The cost function can now be defined as
\begin{equation}
{\rm Cost}(n,m,k,\mu)=\frac{\# {\rm MVM}}{{P_{\rm acc}}}(n,m,k,\mu)\;,
\label{eq:thecost}
\end{equation}
where $\#$MVM is the average number of matrix-vector multiplications involving the operators $Q$ and
$D_2$ over a trajectory. We are neglecting here the cost of computing the gauge contribution
to the driving force, but we checked in a few cases that this is at the few percent level.
It is easy to separate the dependencies on the integrator parameters and on the algorithmic ones (just $\mu$ in our case)
in both $ P_{\rm acc}$ (or better ${\rm Var}(\Delta H)$, as in eq.~\ref{eq:pacc}) and $\#$MVM. Namely, using eq.~\ref{eq:calF}
\begin{eqnarray}
\label{eq:therealthing}
\!\!\!\!\!\!\!{\rm Var}(\Delta H)\!\!\!\! &=& \!\!\! 2 {\rm Var}(\delta H) \! = \!\frac{2\delta \tau^4}{(72)^2}\!\left[ 
{\rm Var}(   \left|{\mathcal{F}}_1\right|^2)(\mu) \!+ \!\frac{ {\rm Var}( \left|{\mathcal{F}}_2\right|^2)(\mu)  }{(4m^2)^2}\!+
\!\frac{ {\rm Var}( \left|{\mathcal{F}}_{\rm G}\right|^2)  }{(16m^2k^2)^2}
 \right],  \\[10pt]
\!\!\!\!\!\!\!\#{\rm MVM} \!\!\! &=& \!\!\! (2n+1) \#{\rm MVM}_1(\mu)+2n(2m+1)  \#{\rm MVM}_2(\mu) \;,
\label{eq:MVM}
\end{eqnarray}
where the index $i$ in  ${\rm MVM}_i(\mu)$ is associated to the operator index ($D_1$ or $D_2$),
or more precisely to the index of the force contribution.
The computation of those requires inverting\footnote{We use the Quasi-Minimal-Residue algorithm for
the inversions in the molecular dynamics. We set the tolerance to $10^{-14}$ in terms of the ratio
between the squared norm of the residue vector and the squared norm of the inversion-source vector.}
 either $Q$ (for ${\mathcal{F}}_1$) or $D_2$ (for ${\mathcal{F}}_2$), see
eq.~\ref{eq:SGS1S2}.
We tried to emphasize in the formulae above that $\mu$ controls the variance of the fermionic forces
and the number of matrix-vector multiplications. We will model those dependencies in an empirical
way and basically fit them. The dependencies on $n,m$ and $k$ are instead completely explicit.

In the definition of the cost we neglected, as customary,  autocorrelations.
Those are difficult to estimate and observable dependent. In addition they
are not strongly affected by algorithmic parameters at fixed trajectory length
and large acceptance, once the form of the HMC preconditioning is chosen~\cite{Meyer:2006ty,DellaMorte:2008ad}.

The proposed optimization strategy should now be clear and to a good extent,
if we were simply aiming at maximizing the acceptance,
it amounts to minimizing \textcolor{black}{ a combination of } 
the variances of the forces as a function of the algorithmic parameters.
We are therefore using the shadow Hamiltonian not only to optimize
the integrators, as previously done~\cite{Clark:2011ir,Kennedy:2012gk}, but also in order to tune 
the parameters defining the factorization of the quark determinant.
\textcolor{black}{As a final remark, one could adopt the same strategy employing better, higher order, integrators,
in which case one should in principle measure the variance of a number of Poisson brackets and miminize 
the cost using the variance of $\delta H$ to estimate the acceptance rate, through a formula equivalent to eq.~\ref{eq:therealthing}.}
%
\section{Tests and Results}
We first test some properties of the integrator used, mostly concerning the size and scaling in $\delta \tau$ 
of the Hamiltonian violations and of the higher order terms in the expansion of the shadow Hamiltonian.
We also compare predictions from different versions of the Creutz formula with actual \textcolor{black}{measurements} of the acceptance.
Finally, after having tested the setup and the accuracy of the approximations introduced, we 
discuss the actual implementation of our optimization procedure and 
present results in the form of predictions and checks.

As already mentioned, we work with two degenerate flavors of (un-improved) Wilson fermions and the plaquette (SU(2)) gauge action.
That is then completely specified by the bare mass parameter $m_0$ and the inverse gauge coupling $\beta=4/g_0^2$.
We use the setup and the code described in~\cite{Hirep} with only a few modifications in order to extract the relevant quantities.
\subsection{Testing properties of the integrator}
\begin{figure}[h!t]
\begin{center}
\includegraphics[scale=0.72]{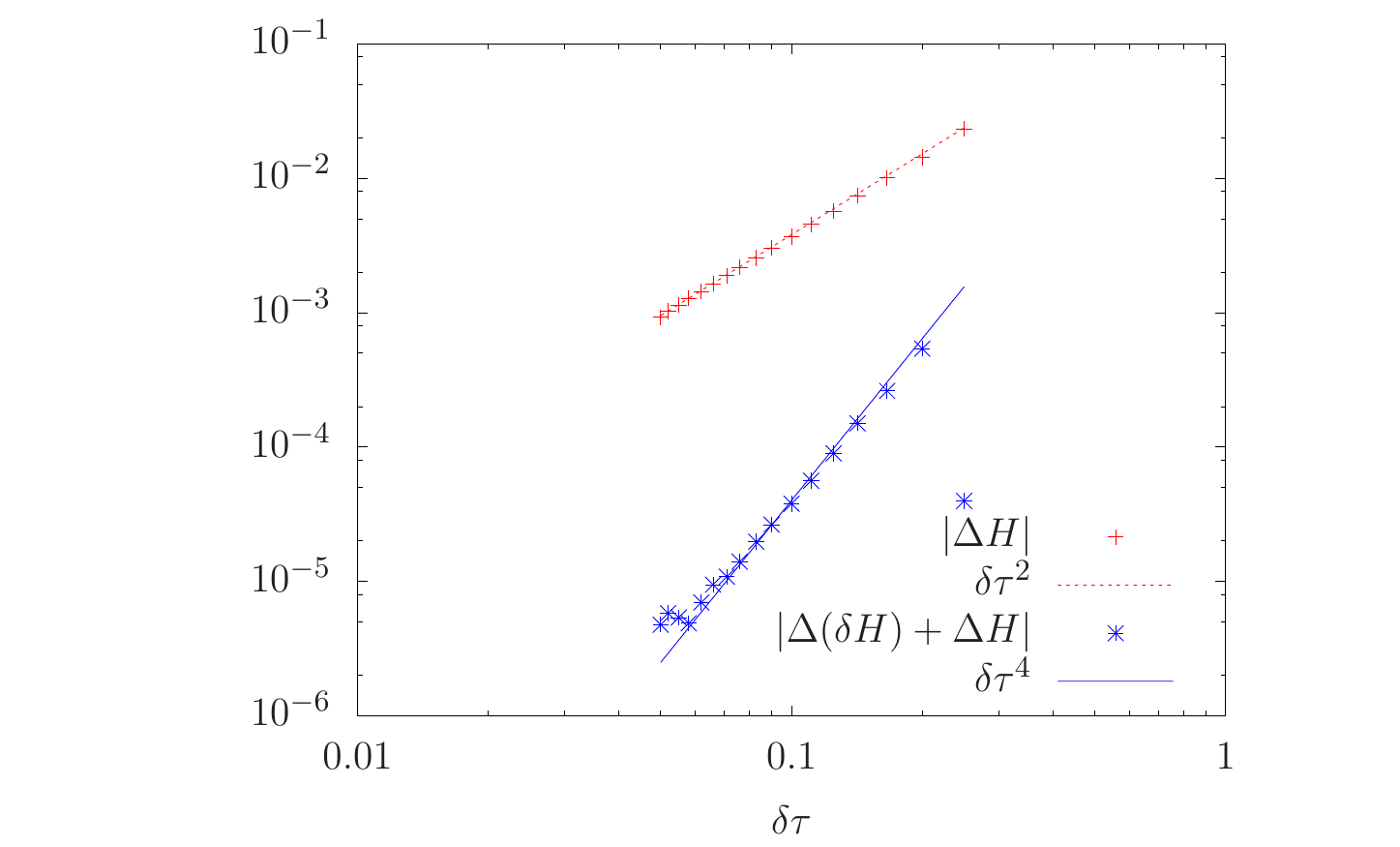}
\caption{Scaling of $|\Delta H|$ and $|\Delta\tilde{H}|=|\Delta H +\Delta(\delta H)|$
with $\delta \tau$ at fixed $\mu=0.2$, $m=10$ and $k=10$.}
\label{fig:Hamviol}
\end{center}
\end{figure}

\begin{figure}[h!t]
\begin{center}
\hspace{-1.cm}
{\subfigure{\includegraphics[scale=0.56]{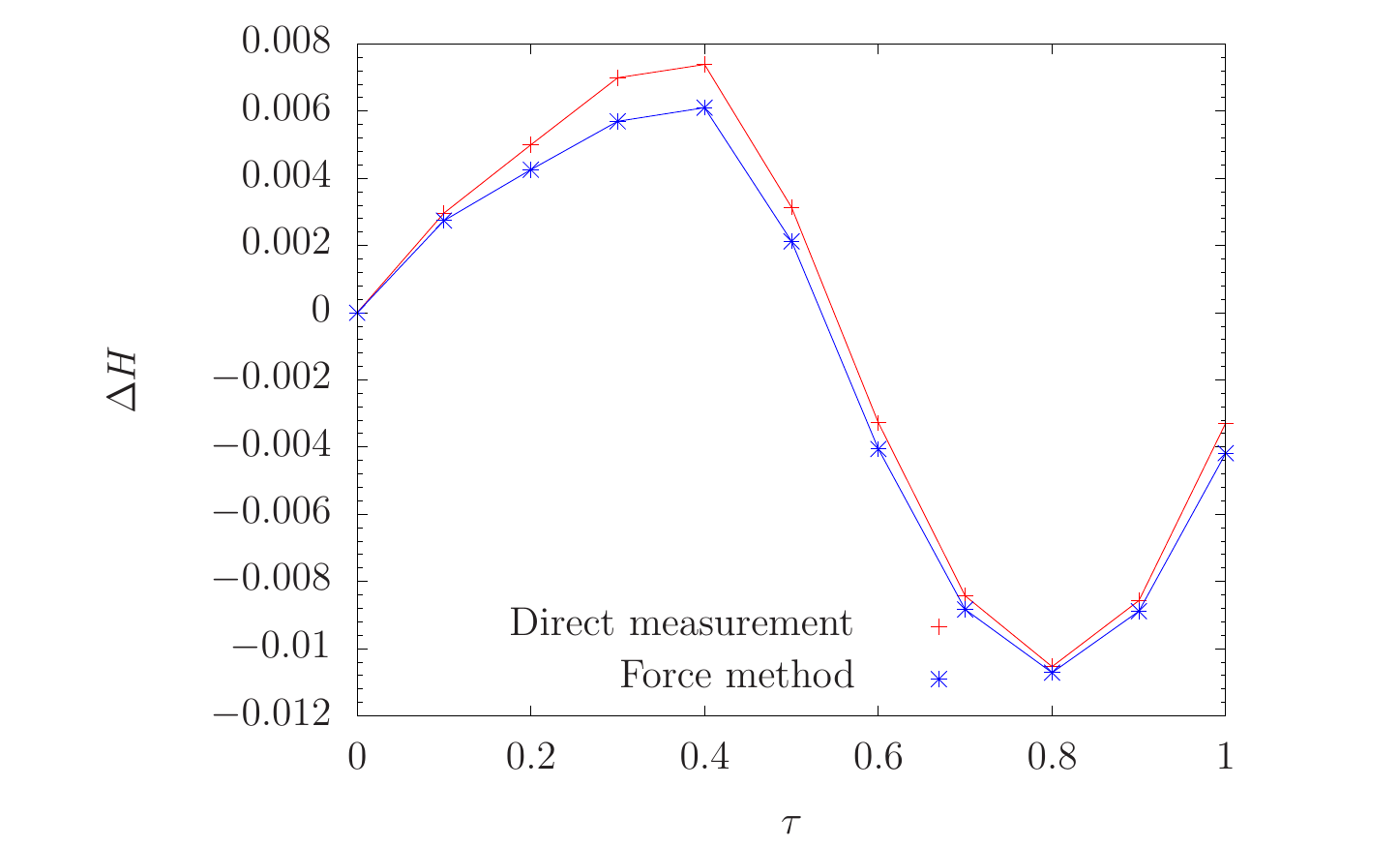}}
\hspace{-1.cm}
\subfigure{\includegraphics[scale=0.56]{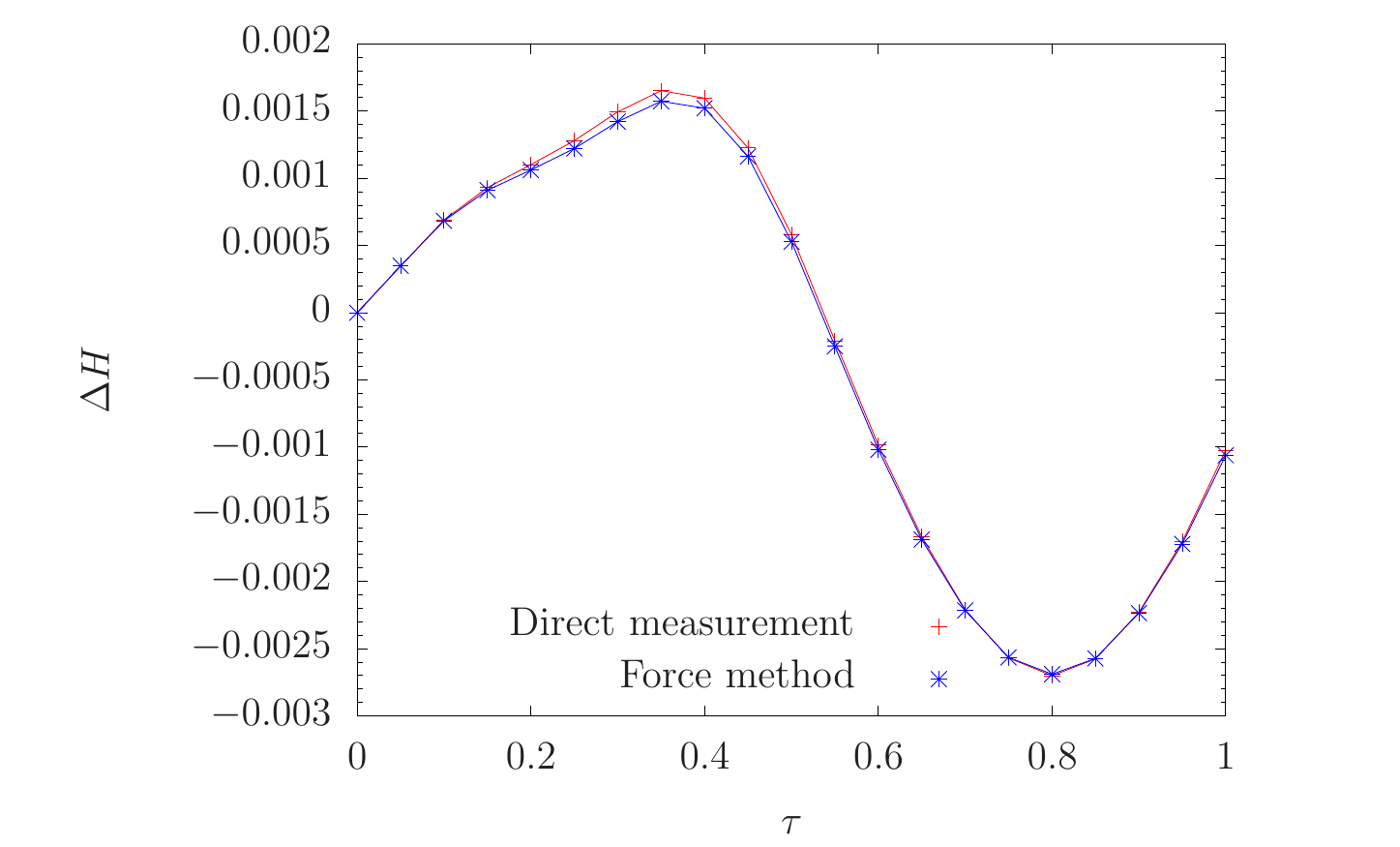}}
\subfigure{\includegraphics[scale=0.8]{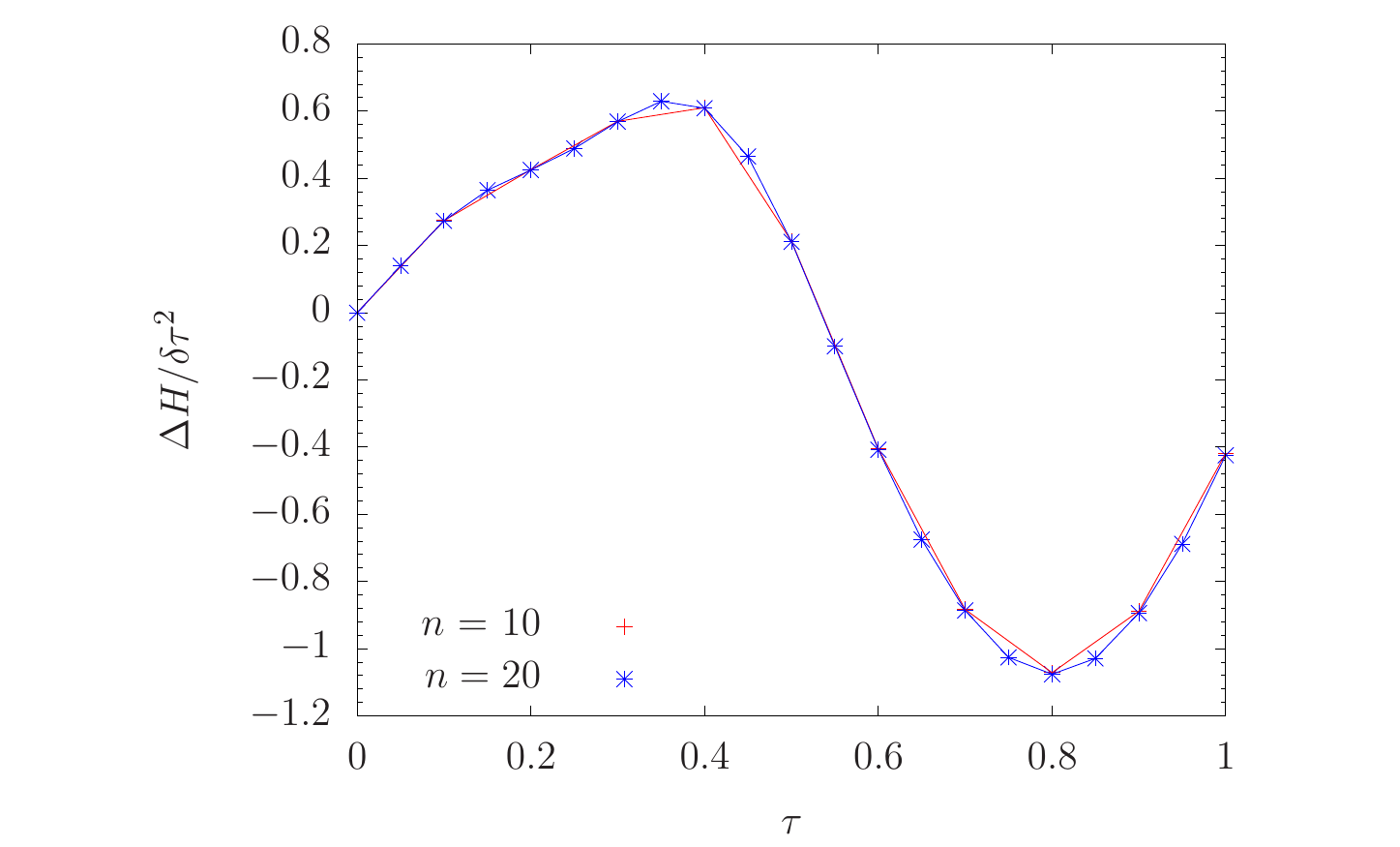}}
}
\caption{Comparison of the Hamiltonian violations computed directly and using the r.h.s. of 
eq.~\ref{eq:niceeq} as a function of the trajectory length and for different time-steps (upper-left:
$\delta \tau=1/10$, $m=10$, $k=10$, upper-right: $\delta \tau=1/20$, $m=10$, $k=10$).
The lower panel shows the data in the upper ones, for the ``Force method'' (i.e., r.h.s. of eq.~\ref{eq:niceeq})  
rescaled by $\delta \tau^2$). In all cases $V=16^4$, $\beta =2.2$, $m_0=-0.72$ and $\mu=0.2$.
}
\label{fig:Hamevol}
\end{center}
\end{figure}

A first simple test of the integrator concerns the scaling of $|\Delta H |$
as a function of $\delta \tau$ at constant trajectory length. Given that we are using a second order
integrator, such Hamiltonian violations should be  $O(\delta \tau^2)$.
\textcolor{black}{In order to study that, we fix the 
the initial configurations of gauge links and conjugate momenta 
and run a trajectory of length one using different values of $\delta \tau$.
In that way we obtain different discretized solutions of the same 
continuum  Hamiltonian problem.}
Similarly one can look at $|\Delta \tilde{H} |$. If we were able to compute
the shadow Hamiltonian exactly, that would clearly vanish, however, since we can
only compute $\tilde{H}$ to a given order in $\delta \tau$ (by computing Poisson brackets),
the shadow Hamiltonian violations should scale as the first neglected term. In
our case, using eq.~\ref{eq:calF}, that means as $\delta \tau^4$.
That is precisely what one concludes from the data plotted in Fig.~\ref{fig:Hamviol}, which 
have been obtained by evolving a $8^4$ configuration at $\beta =2.2$ and $m_0=-0.72$ over
one trajectory using different time-steps.
It is also instructive to look at the scaling of $\Delta H$ with the trajectory length $\tau$.
The conservation of the shadow Hamiltonian implies $\Delta H = -\Delta (\delta H)$, and again,
by rewriting $\delta H$ using eq.~\ref{eq:calF} one gets
\begin{equation}
\Delta H = -
\frac{\delta \tau^2}{72} \Delta \left( \left|{\mathcal{F}}_1\right|^2 + 
\frac{ \left|{\mathcal{F}}_2\right|^2 }{4m^2}  + \frac{ \left|{\mathcal{F}}_{\rm G}\right|^2 }{16m^2k^2}      \right)\;,
\label{eq:niceeq}
\end{equation}
up to $O(\delta \tau^4)$. For one given, thermalized, initial configuration, the quantities on the two sides 
of the above equation are compared as a function
of $\tau$ for two different values of $\delta\tau$ in the two upper panels of Fig.~\ref{fig:Hamevol}, while in the
lower panel the quantity $\Delta H/\delta \tau^2$ is plotted. As suggested by eq.~\ref{eq:niceeq} such ratio is 
independent from the step-size (up to $O(\delta \tau^2)$), since the same underlying continuous Hamilton equations
with the same initial conditions are being discretized in the two cases.
\subsection{Testing estimates of the acceptance}
The previous tests serve the purpose of checking our computation 
of $\delta H$ up to $O(\delta \tau^4)$. Here we want to assess how robust the
estimates of the acceptance using $\delta H$  are. 
It is by modeling such expressions that we will be able to predict
the efficiency of a simulation for a given choice of the parameters $\mu,\, n,\, m$ and $k$.

\begin{figure}[h!t]
\begin{center}
{\subfigure{\includegraphics[scale=0.53]{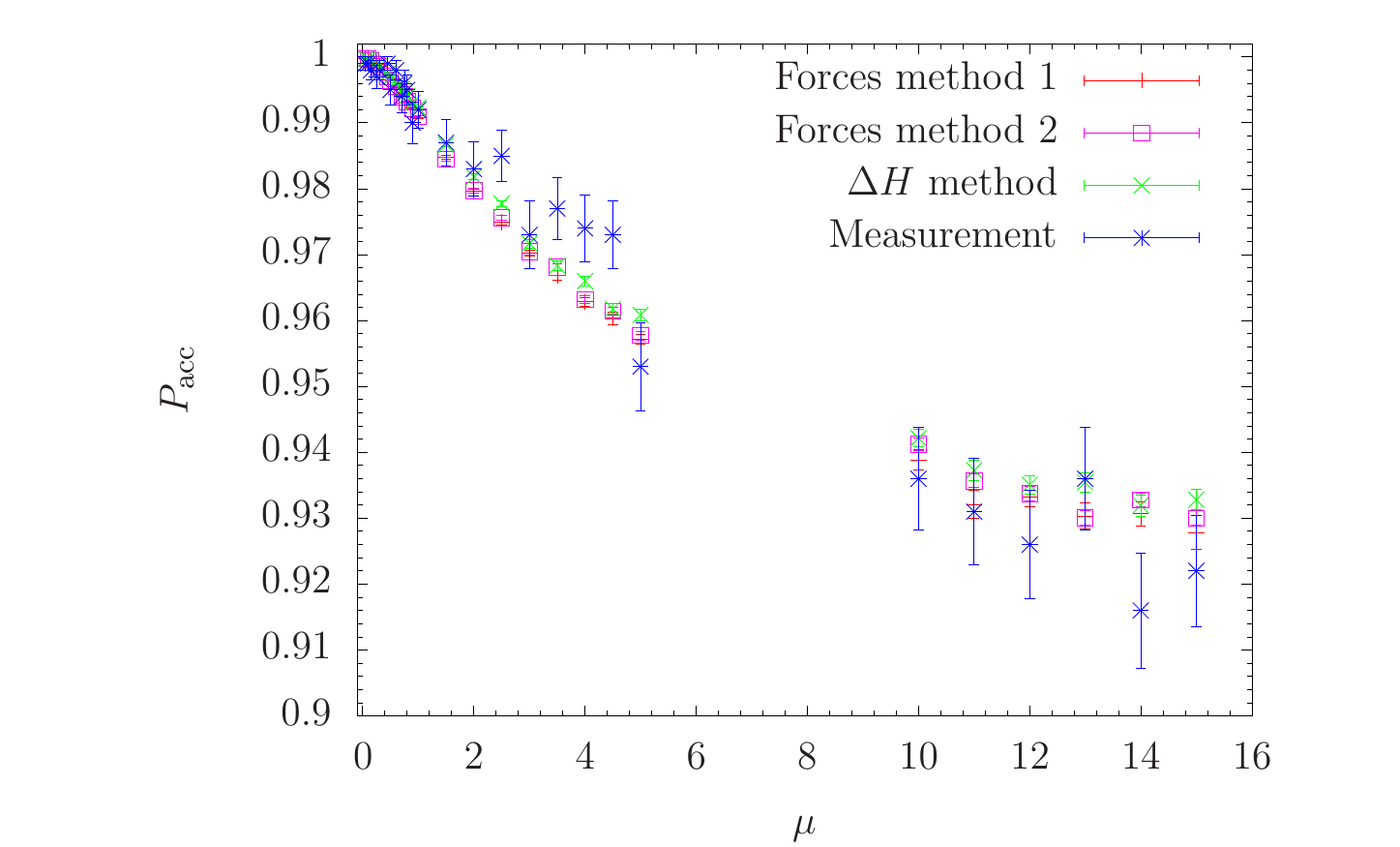}}
\hspace{-1.cm}
\subfigure{\includegraphics[scale=0.53]{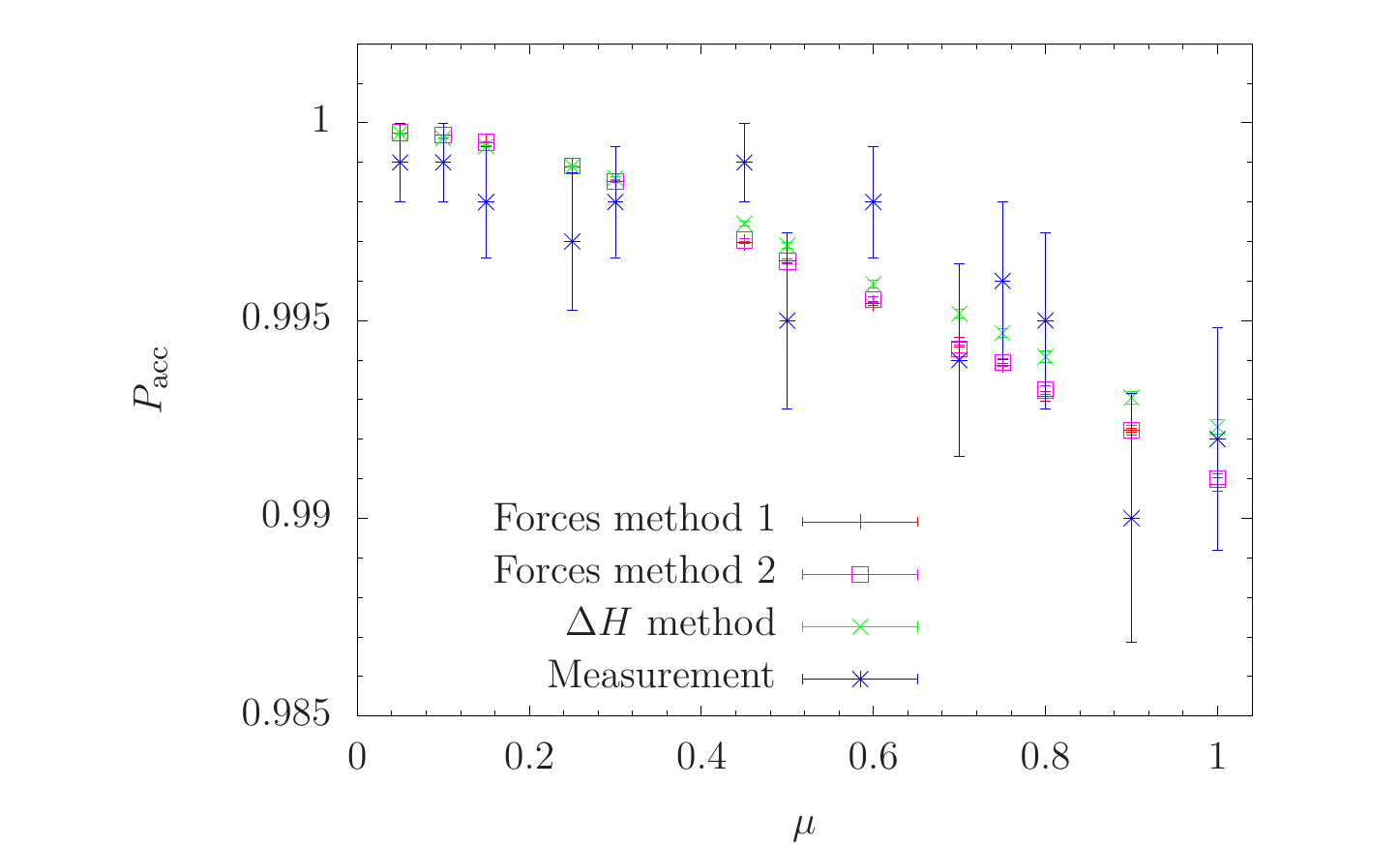}}
}
\caption{
Comparison of the four different estimates of the acceptance described in the
text as a function of the parameter $\mu$. Results refer to a $4^4$ lattice
at  $\beta =2.2$ and $m_0=-0.72$. We fixed $n=4$,  $m=k=10$ and we ran
$O(10^3)$ trajectories of length 1 at each value of $\mu$.
\textcolor{black}{On the left the full range of $\mu$ explored, on the right a zoom of the area $0<\mu \leq 1$.}
}
\label{fig:acceptances}
\end{center}
\end{figure}

We can compare four different estimates of the acceptance:
\begin{itemize}
\item[$i$] is obtained by trivially counting the number of accepted configurations ('Measurement'
in Fig.~\ref{fig:acceptances}),
\item[$ii$] requires measuring ${\rm Var}(\Delta H)$, which is then plugged into eq.~\ref{eq:pacc} 
('$\Delta H$ method' in Fig.~\ref{fig:acceptances}),
\item[$iii$] instead of $\Delta H$ one measures the quantity on the r.h.s.~of eq.~\ref{eq:niceeq} and 
its variance, which is then inserted in eq.~\ref{eq:pacc} replacing  ${\rm Var}(\Delta H)$ 
('Forces method 2' in Fig.~\ref{fig:acceptances}),
\item[$iv$] similarly to the estimate in $iii$, one replaces ${\rm Var}(\Delta H)$, in eq.~\ref{eq:pacc}
with the r.h.s. of eq.~\ref{eq:therealthing}. Notice that in this case the variance is computed
not only over the trajectories (as in $ii$ and $iii$), but also over the inner trajectory steps.
This estimate ('Forces method 1' in Fig.~\ref{fig:acceptances}),
is therefore expected to be the most reliable.
\end{itemize}
%
From Fig.~\ref{fig:acceptances}
one concludes that, for large values of $P_{\rm acc}$, all estimates agree quite well 
with the measured acceptance within its errors.
Also, they vary in a very smooth way, which suggests that the estimates are quite stable, and
$O(10^3)$ trajectories (as used for the figure) are enough to obtain reliable values.
In the following we will use the estimate $iv$ from the list above.
\subsection{Results and modelling of the data}
\begin{figure}[htb]
\begin{center}
\includegraphics[scale=0.67]{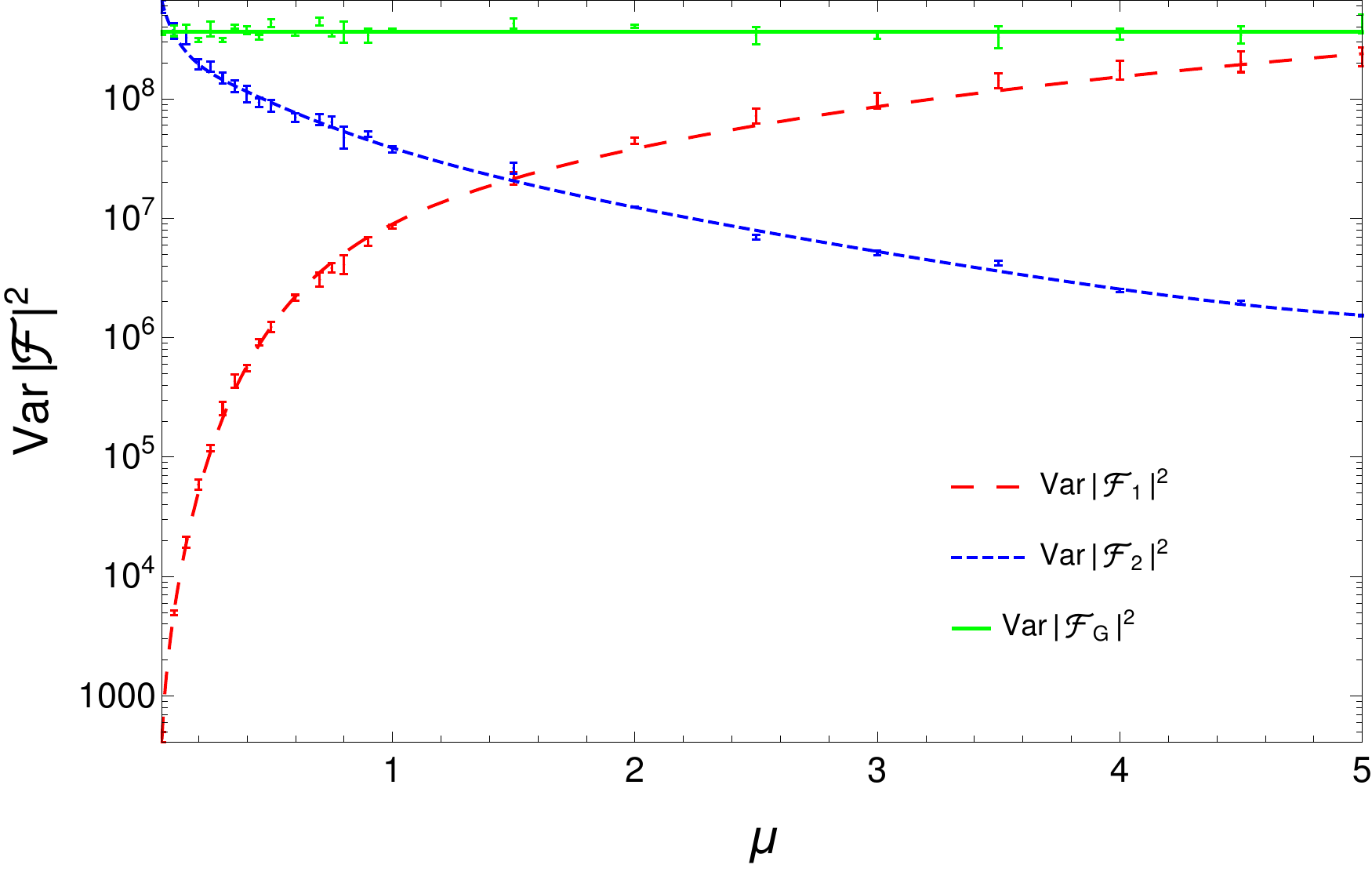}
\caption{Variances of the forces as a function of the mass preconditioning parameter from a set of simulations
at $\beta=2.2$, $m_0=-0.72$ and $V=32^4$. Curves only serve to guide the eye.
}
\label{fig:variances}
\end{center}
\end{figure}

The key observation in the proposed approach is that the variances in eq.~\ref{eq:therealthing} are functions of $\mu$ only,
at fixed physical parameters (quark masses, gauge coupling and volume).
We plot such variances as a function of the Hasenbusch mass-parameter $\mu$ 
for a $32^4$ lattice at $\beta=2.2$ and $m_0=-0.72$ in 
Fig.~\ref{fig:variances}. These results have been obtained for $n=15$, $m=8$ and $k=10$ fixed,
but in principle we could have combined results for different choices of such algorithmic parameters,
as they should not affect the variances.
One can see that the variance of the gauge force is independent from $\mu$, as expected, whereas
for ${\cal{F}}_1$ and  ${\cal{F}}_2$ a region of weak dependence, for large values of $\mu$ and 
a region of strong dependence for small ones can be clearly identified. The interesting region
for us is the one at small $\mu$, since there the hierarchy of the forces is consistent with the nested
scheme adopted for the integrator (see remarks in Sect.~\ref{onHMC}).

\begin{figure}[h!t]
\hspace{-1.18cm}
{\subfigure{\includegraphics[scale=0.4]{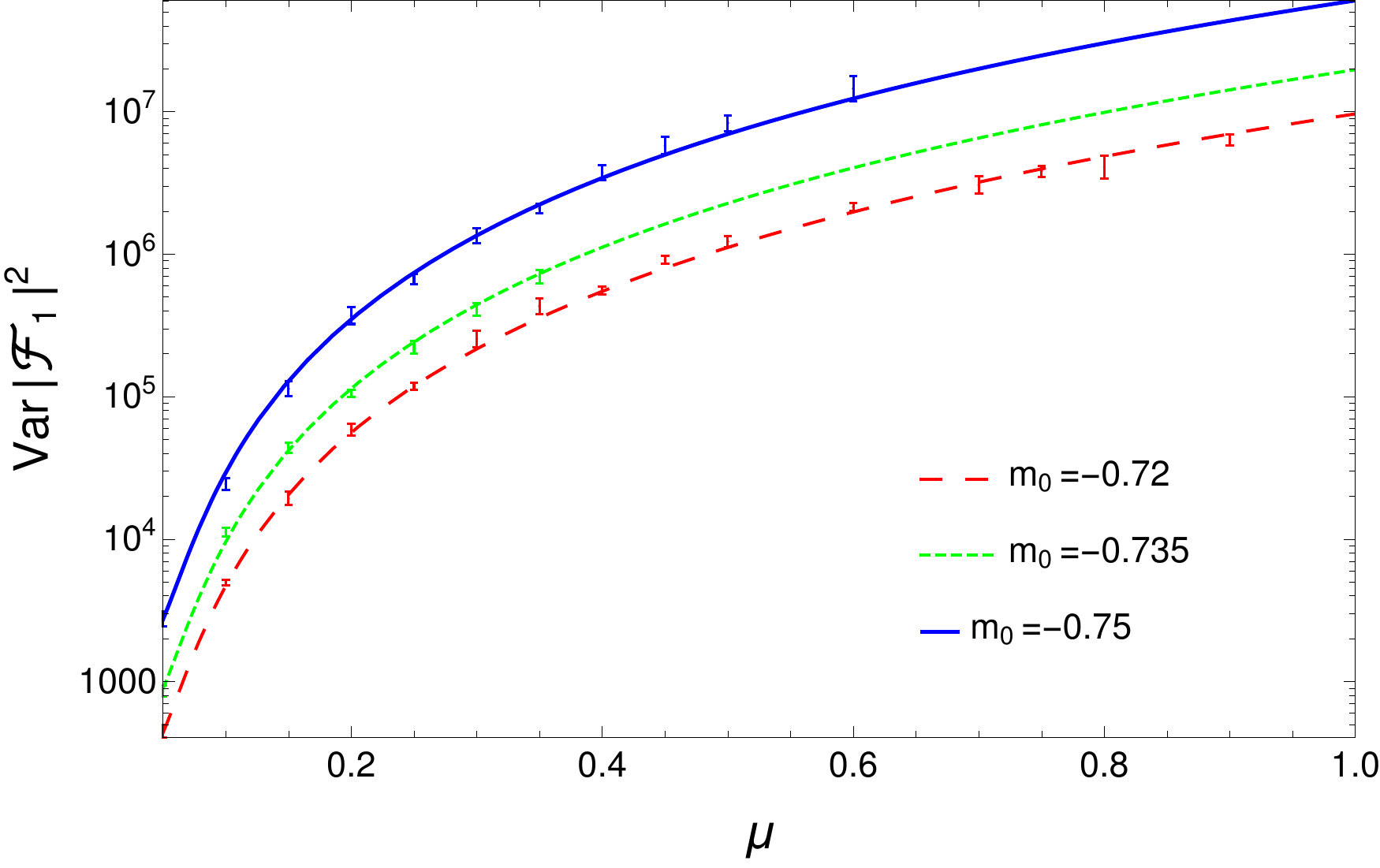}}
\hspace{-1.5cm}
\subfigure{\includegraphics[scale=0.4]{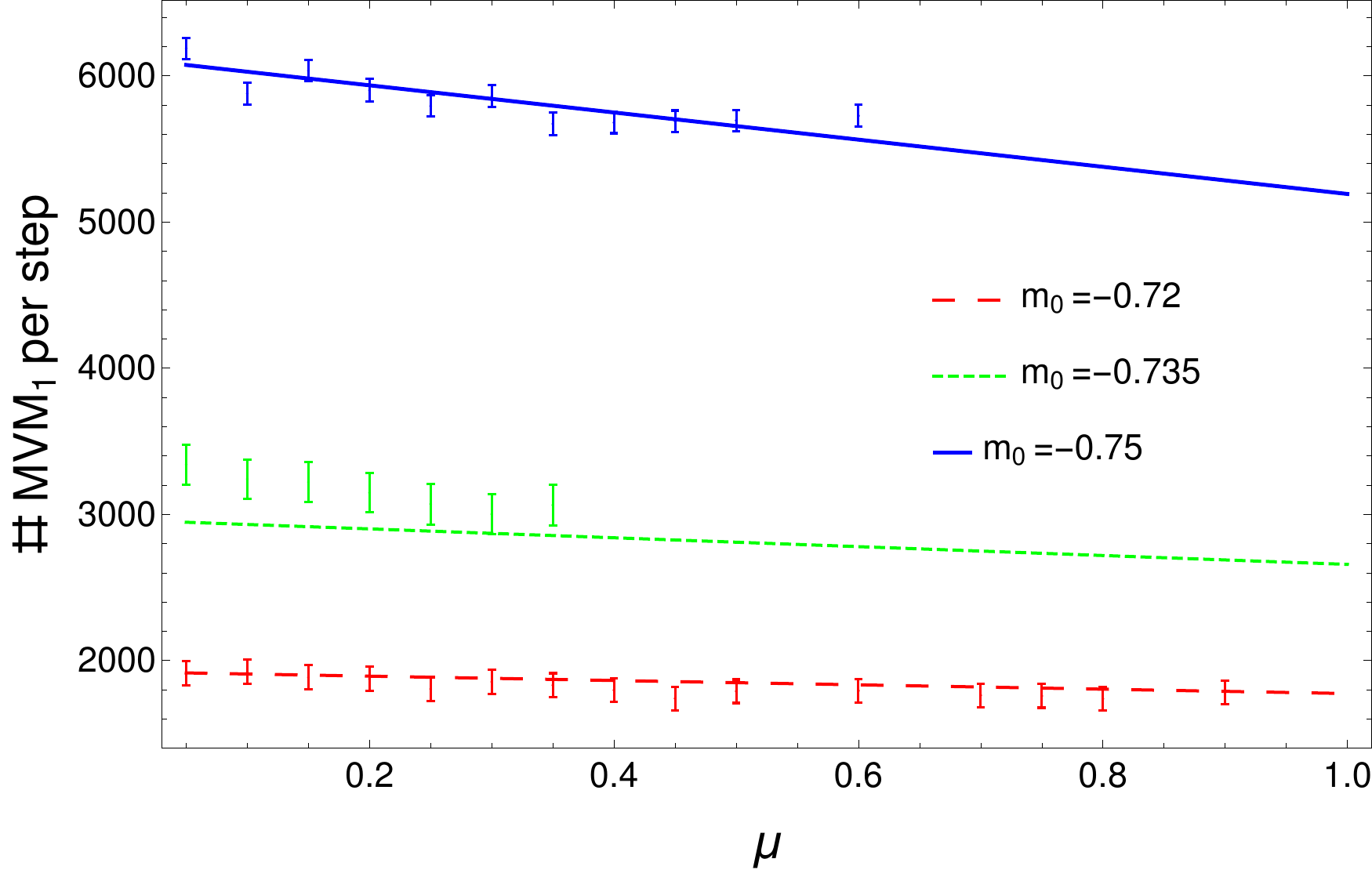}}
\hspace{-1.8cm}
\subfigure{\includegraphics[scale=0.4]{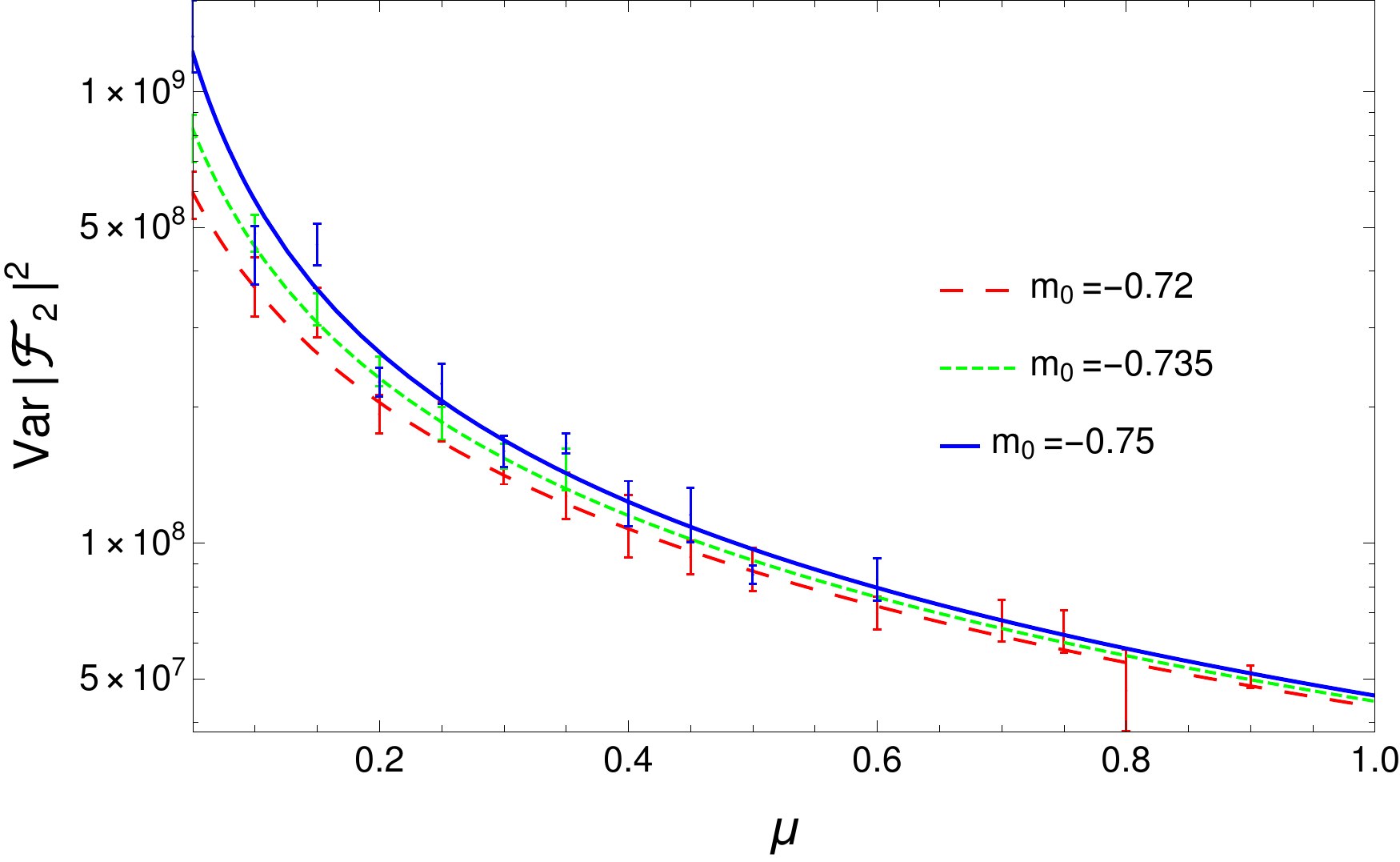}}
\hspace{-1.8cm}
\subfigure{\includegraphics[scale=0.4]{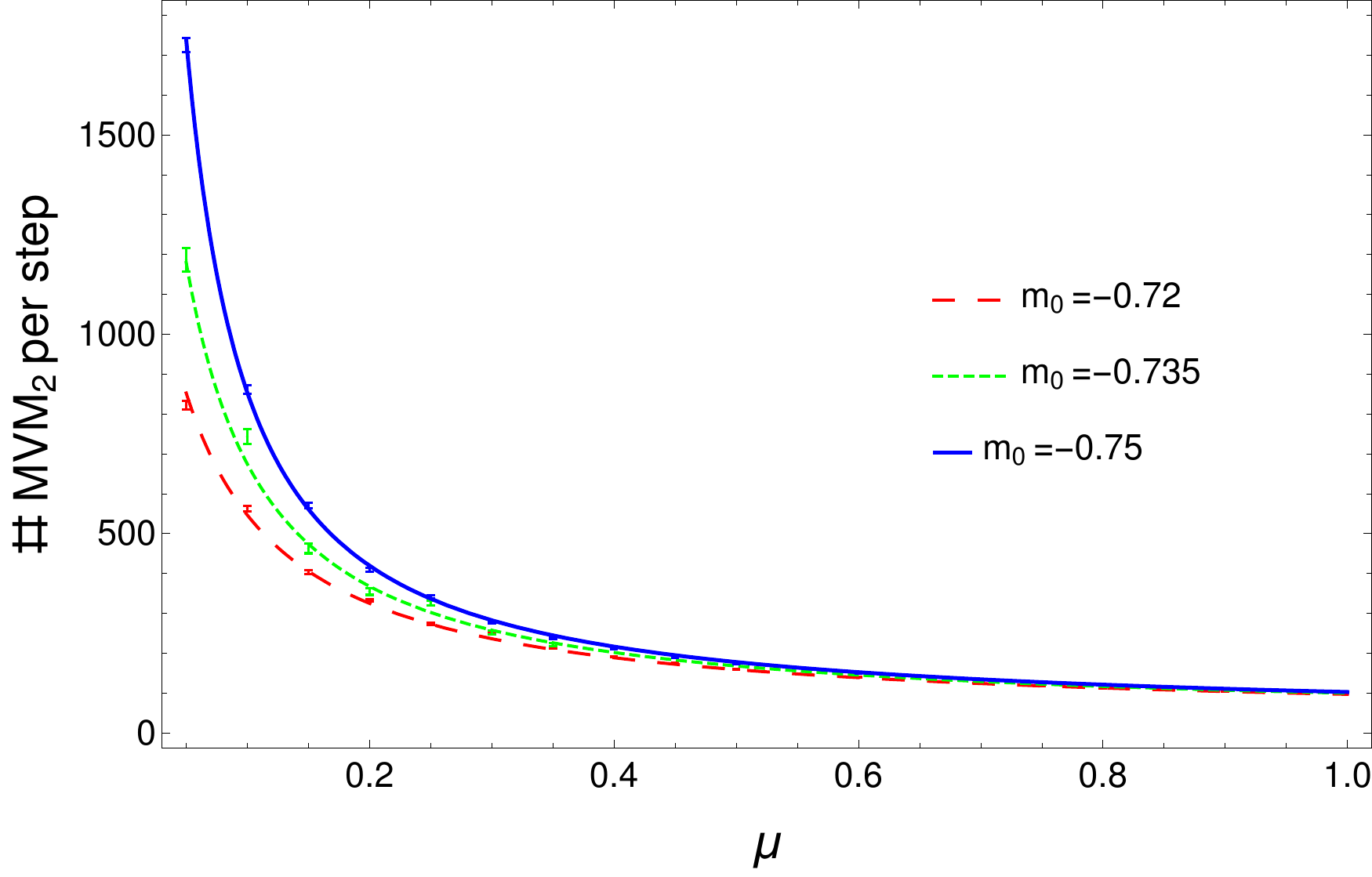}}
\hspace{-4.cm}
\subfigure{\includegraphics[scale=0.4]{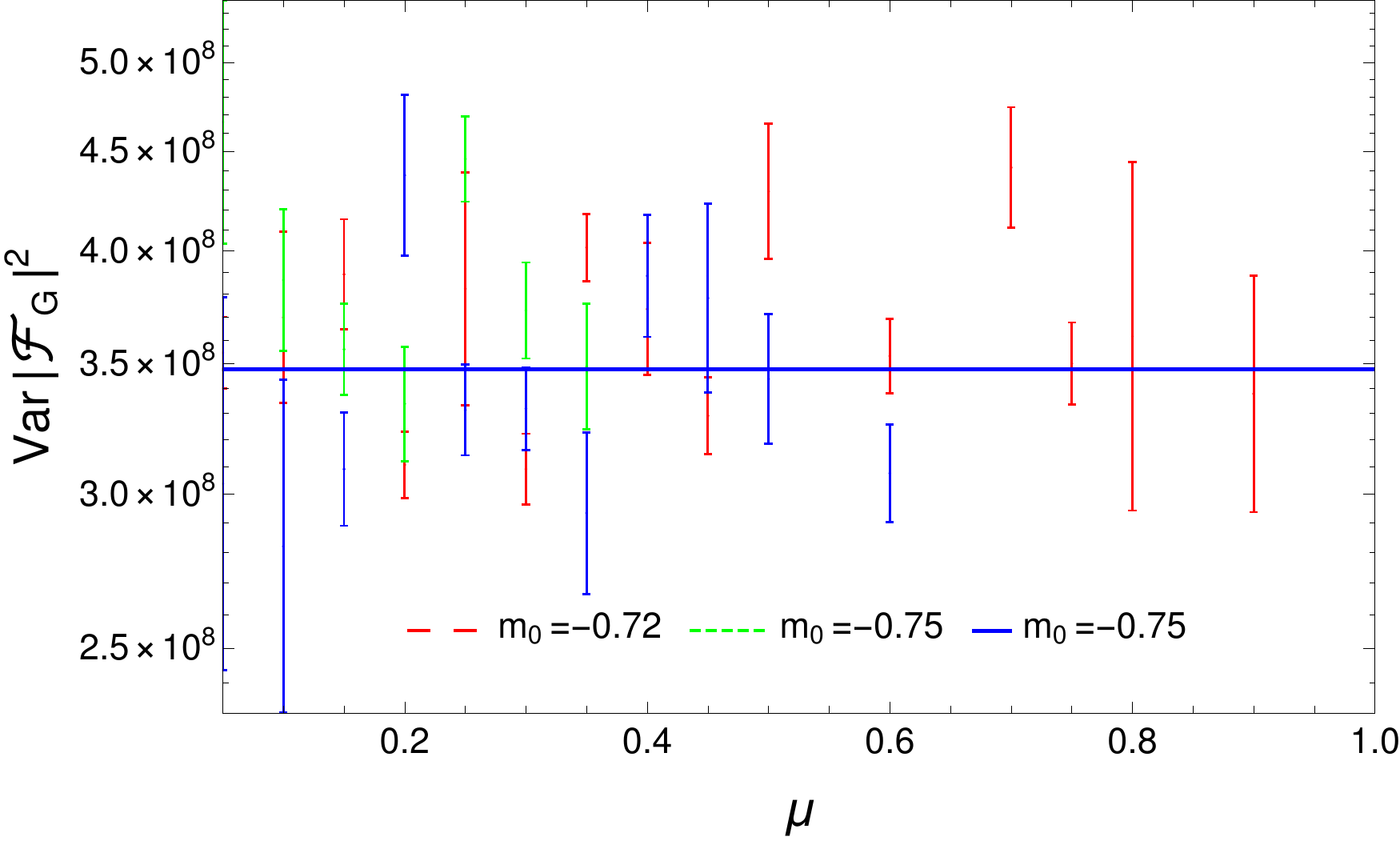}}
}
\caption{Left column: global fits to the forces' variances as a function of $\mu$ for three different quark masses.
Right column: global fits to the average number of matrix-vector multiplications per step in the computation of the fermionic forces vs $\mu$
and for $m_0=-0.72$, $-0.735$ and $-0.75$. The inverse gauge coupling $\beta$ is fixed to $2.2$ and $V=32^4$.
} 
\label{fig:lotoffits}
\end{figure}

Although a precise theoretical prediction on the functional form describing the dependence of such
variances on $\mu$ is lacking, for small values of $\mu$ (say $\mu\lesssim 1$) one can use simple
polynomial fits and impose a few reasonable constraints as the vanishing of
 ${\rm Var}(   \left|{\mathcal{F}}_1\right|^2)$ for $\mu=0$ and, following~\cite{DelDebbio:2005qa},  a rapid growth of 
${\rm Var}(   \left|{\mathcal{F}}_2\right|^2)$  for values of $\mu$ compensating for the quark mass 
(i.e., $\mu \simeq -(m_0-m_{\rm c})$, with $m_{\rm c}$ the critical value of the bare quark mass).
Similar considerations can be used for the fit {\it{ans\"atze}} of the $\#{\rm MVM}_i(\mu)$ functions.
We therefore adopt the following parameterizations:
\begin{eqnarray}
{\rm Var}(   \left|{\mathcal{F}}_1\right|^2)(\mu) &=& a \mu + b\mu^2 +c\mu^3\;,\\
{\rm Var}(   \left|{\mathcal{F}}_2\right|^2)(\mu) &=& \frac{a + b\mu +c\mu^2}{(\mu + m_0 - m_{\rm c})^2} \;, \\
{\rm Var}(   \left|{\mathcal{F}}_{\rm G}\right|^2)(\mu) &=& a\;, \\
\#{\rm MVM}_1(\mu) &=& a +b \mu\;, \\
\#{\rm MVM}_2(\mu) &=& \frac{a + b\mu +c\mu^2}{(\mu + m_0 - m_{\rm c})^2} \;, 
\end{eqnarray}
where of course the fit parameters $a,\, b$ and $c$ are independent for each of the quantities above.
Notice that we include both even and odd terms in $\mu$ since the splitting we use is not invariant under
$\mu \leftrightarrow -\mu$ (see  Sect.~\ref{onHMC}).
The fits provide good descriptions of the data with reasonable $\chi^2/d.o.f$ values. We stress however
that any other parameterization reproducing the data would be equally good for the present purposes.
In principle one could even decide to mimimize the algorithmic cost on a finite (fine) set of $\mu$ values,
which would completely bypass the problem of modelling the results through smooth functional forms.
At fixed quark mass, we will compare the two approaches and show that the choice does not significantly
affect the final estimates, which gives us confidence on the chosen parameterizations.
The advantage of modelling the data is that one can simultaneously fit the dependencies
on $\mu$ and on the quark mass, which helps stabilizing the results when different sets of 
$\mu$ values have been considered for different quark masses.
That is precisely our case. While for $m_0=-0.72$ we performed a rather fine scan in $\mu$,
for $m_0=-0.735$ and $m_0=-0.75$ we ran simulations on a reduced set of $\mu$ values only (the 
lattice volume is fixed to $32^4$ at  $\beta=2.2$, where $m_{\rm c}=-0.7676(2)$).
In order to account for the quark mass dependence we promote some of the coefficients in the fit-forms
above to functions of the bare subtracted quark mass and finally use: 
\begin{eqnarray}
{\rm Var}(   \left|{\mathcal{F}}_1\right|^2)(\mu,m_0) &=& \frac{a \mu + b\mu^2 +c\mu^3}{(m_0-m_{\rm c})^2}\;,\\
{\rm Var}(   \left|{\mathcal{F}}_2\right|^2)(\mu,m_0) &=& \frac{a + b\mu +c\mu^2}{(\mu + m_0 - m_{\rm c})^2} \;, \\
{\rm Var}(   \left|{\mathcal{F}}_{\rm G}\right|^2)(\mu,m_0) &=& a\;, \\
\#{\rm MVM}_1(\mu,m_0) &=& \frac{a + b (m_0-m_{\rm c}) + c \mu}{(m_0-m_{\rm c})^2}\;, \\
\#{\rm MVM}_2(\mu,m_0) &=& \frac{a + b\mu +c\mu^2}{(\mu + m_0 - m_{\rm c})^2} \;.
\end{eqnarray}
The resulting fits are shown in Fig.~\ref{fig:lotoffits}.
\subsection{Cost predictions and verifications}
Having now smooth functions of $\mu$ describing the variances in eq.~\ref{eq:therealthing}
and the $\#$MVM$_i$ in eq.~\ref{eq:MVM}, we plug the first in eq.~\ref{eq:pacc} and then in
the denominator of the r.h.s. of eq.~\ref{eq:thecost} and the second in the numerator
of the same expression, to obtain the Cost in closed form as a function of $n\;,m\;,k$ 
and $\mu$. At this point we minimize such function using {\tt{Mathematica}}, under the
additional constraint $P_{\rm acc}\gtrsim 0.75$, since for example the Creutz formula we used is valid in
principle for large acceptances only. We further fix $k=10$ in order to reduce the parameter space 
for the minimization. The choice may be conservative, but since $k$ controls the frequency of the
computation of the gauge force, which is very cheap, we do not expect large gains in optimizing
this parameter. 

\begin{figure}[h!t]
\hspace{-1.18cm}
{\subfigure{\includegraphics[scale=0.6]{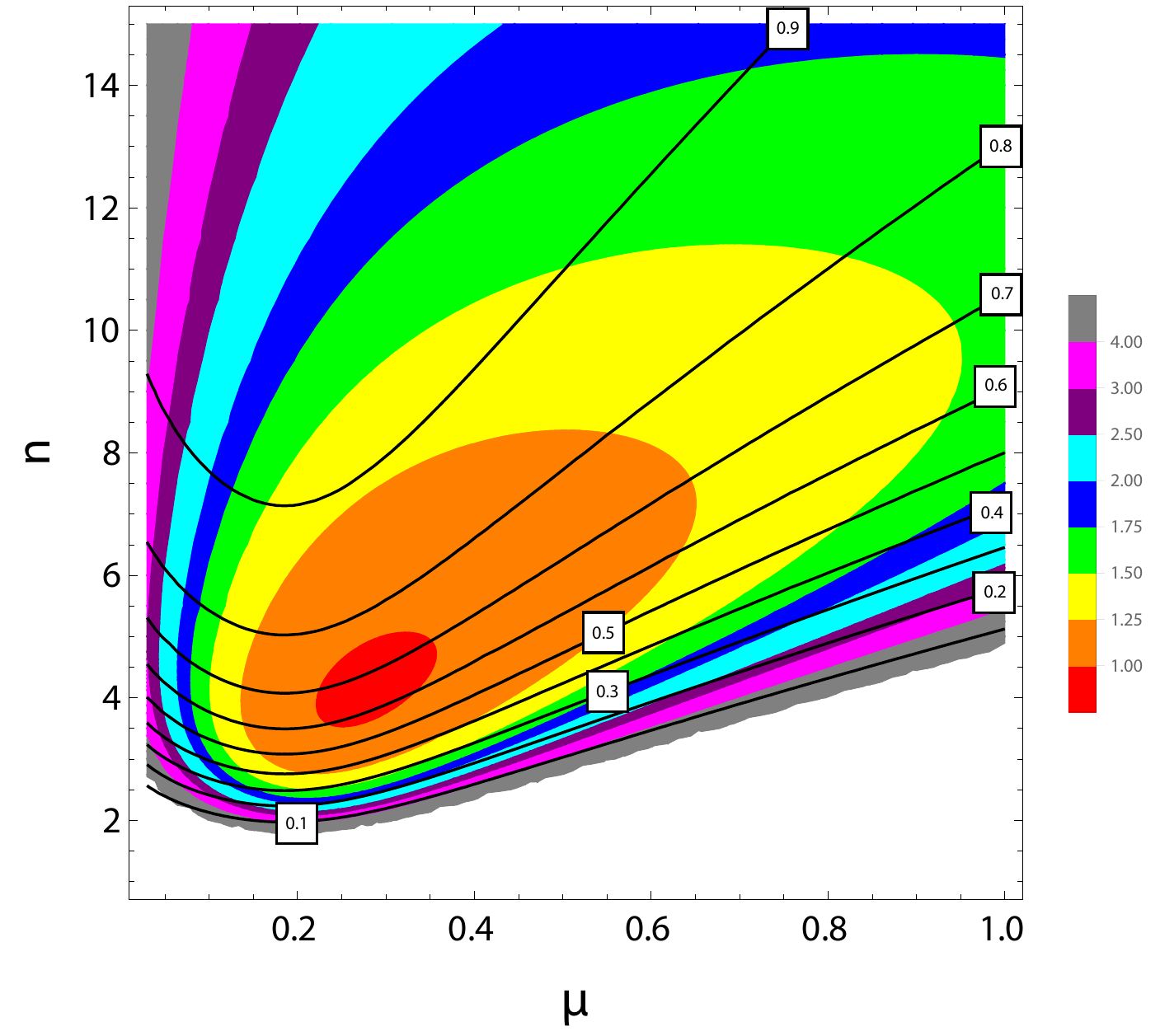}}
\subfigure{\includegraphics[scale=0.6]{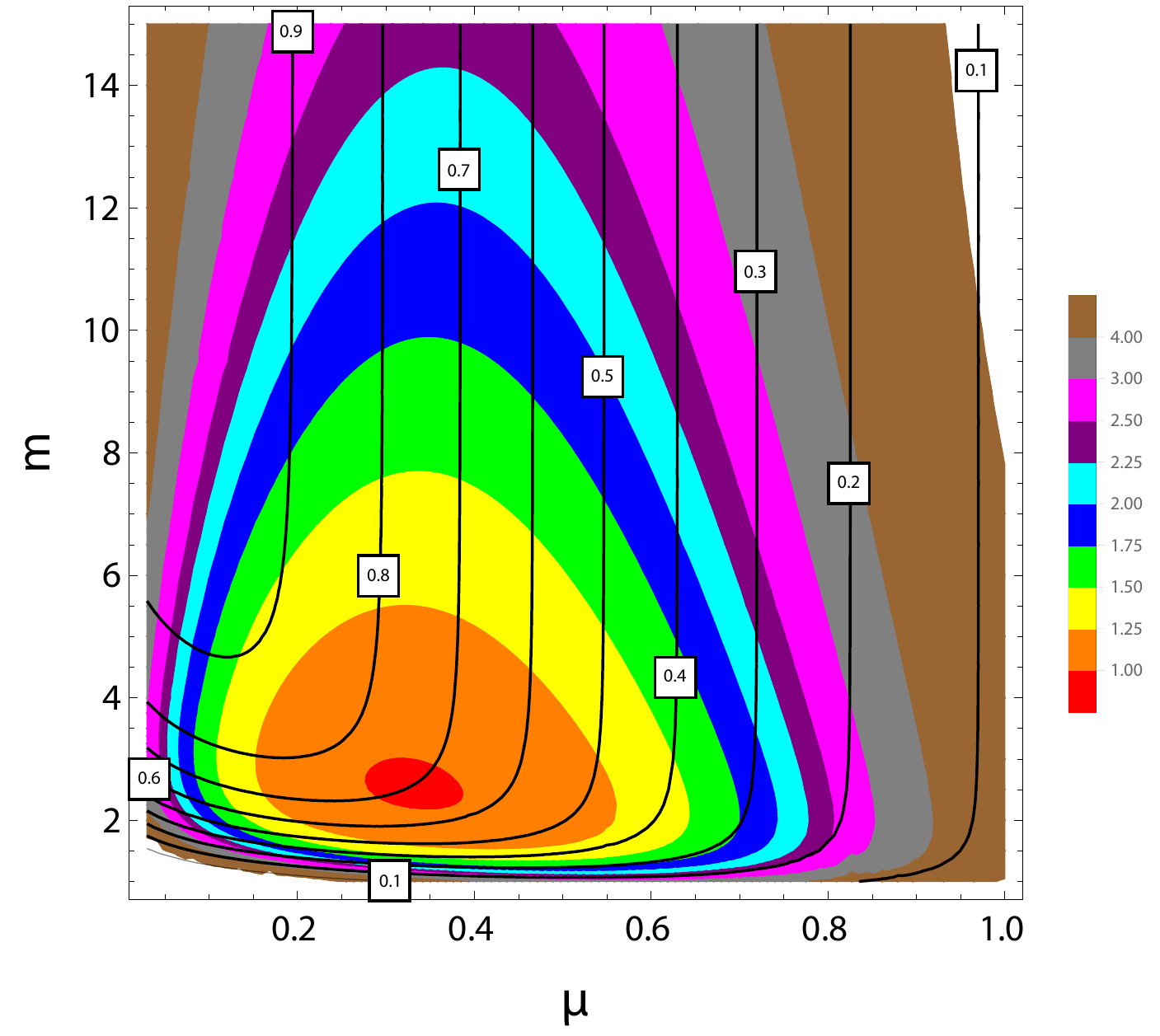}}
}
\caption{Slices of the Cost function at fixed $m=3$ and $k=10$ as a function of $\mu$ and $n$ (left) and
at fixed $n=5$ and $k=10$ as a function of $\mu$ and $m$ (right). Solid lines 
represent constant acceptance curves. The cost is expressed relatively to the minimum.
$\beta=2.2$, $m_0=-0.72$ and $V=32^4$. 
} 
\label{fig:levelcurv}
\end{figure}

What we obtain are level curves as the ones shown in Fig.~\ref{fig:levelcurv}. One can see that the 
minima are rather broad, especially as a function of $\mu$.

\begin{figure}[h!t]
\begin{center}
\includegraphics[scale=0.67]{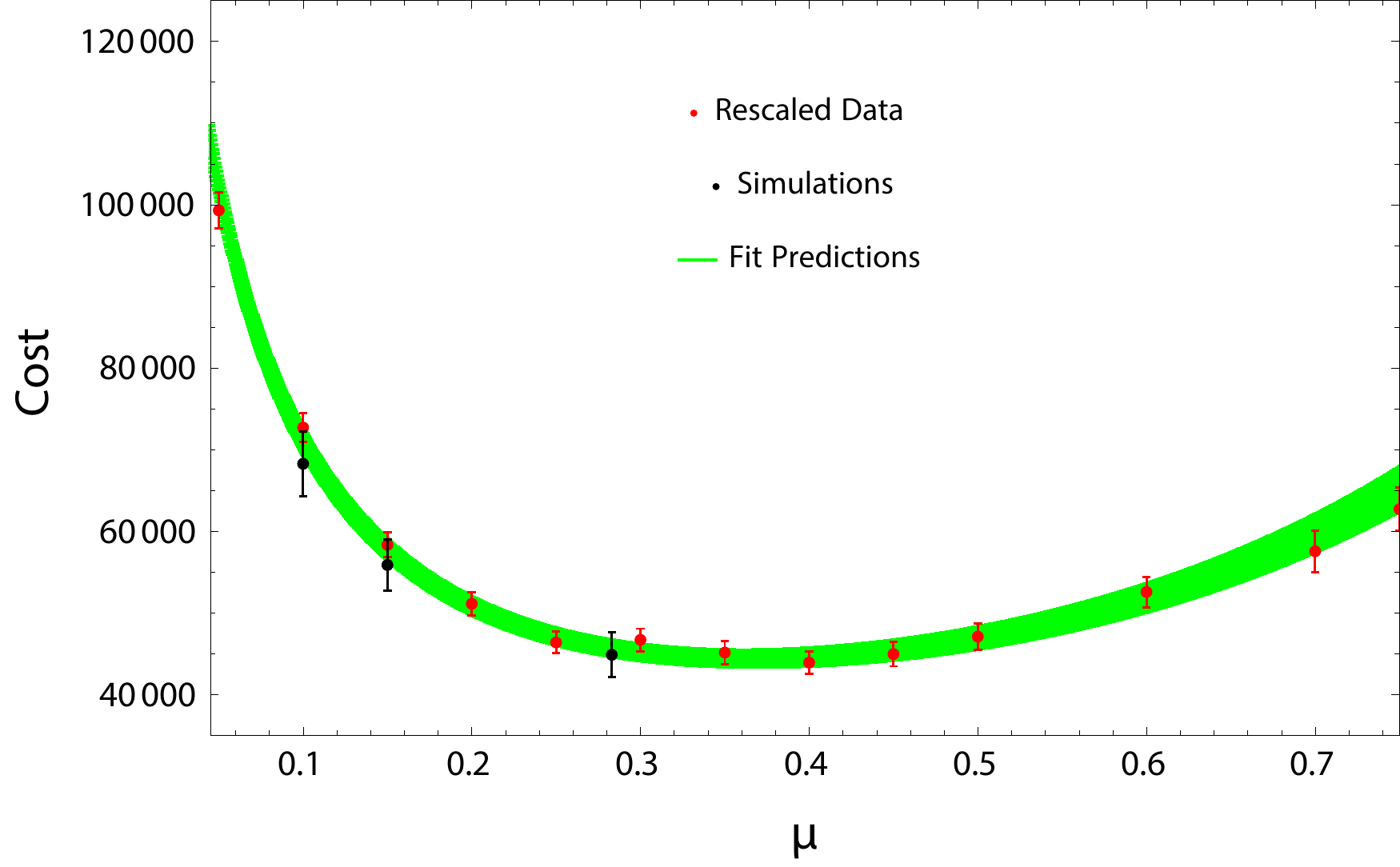}
\caption{Comparison of the Cost between direct simulations and our final predictions.
 For the second we either used measurements at fixed values of $\mu$ (``rescaled data''
in the plot) or we used fit results (``Fit prediction''). 
$n$, $m$ and $k$ are fixed to $5$, $3$ and $10$ respectively (again, $\beta=2.2$, $m_0=-0.72$ and $V=32^4$).
Each simulation consists of about 1000 molecular dynamics units (MDU).
}
\label{fig:costvsmudirect}
\end{center}
\end{figure}

Since we have the full Cost function (not just the minimum), we can compare predictions
to actual simulations. We do that at fixed $n$, $m$ and $k$ (those corresponding
to the minimum for the cost) because we are mostly interested in checking, a-posteriori, the 
assumptions made in fitting the dependencies on $\mu$ as described in the previous Section.
In Fig.~\ref{fig:costvsmudirect} we compare predictions obtained either by using the variances
and average numbers of matrix-vector multiplications measured at specific values of $\mu$
(so without any fitting) or by using the parameterized forms, to actual simulations.
We see that in all cases the agreement is rather good, over a region where the cost changes
by a factor of about three.

\begin{figure}[h!t]
\begin{center}
\includegraphics[scale=0.67]{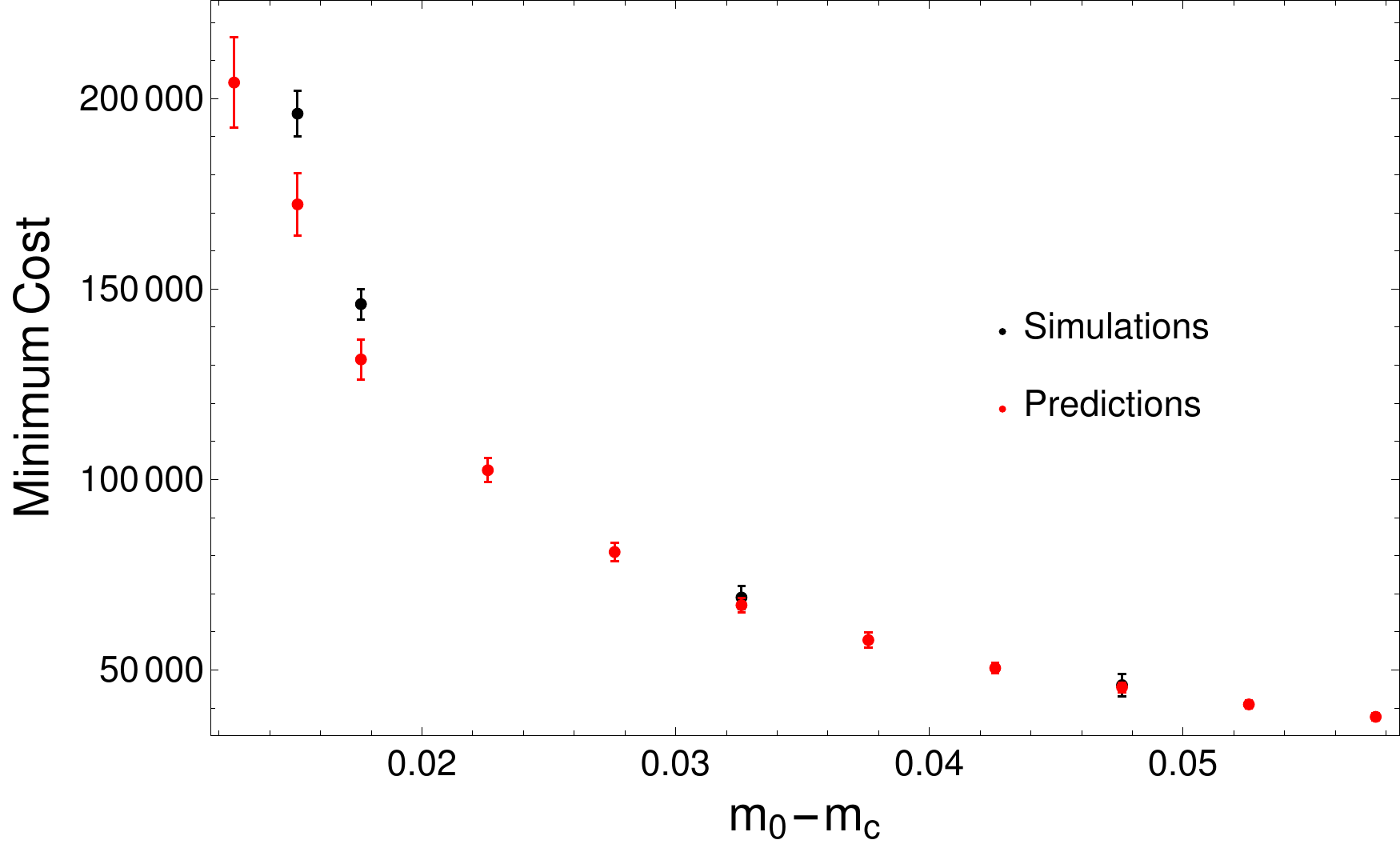}
\caption{Cost from direct simulations and as predicted within our approach
as a function of the bare subtracted quark mass. The algorithmic parameters at each point
are chosen such that the cost is minimized  (as usual $\beta=2.2$ and $V=32^4$).
For each simulation we ran between  500 and 1000 MDU.
}
\label{fig:costvsm0direct}
\end{center}
\end{figure}

Finally, in Fig.~\ref{fig:costvsm0direct} we compare our prediction for the
dependence of the cost on the quark mass, to direct simulations.
In this case we always fix the algorithmic parameters to the obtained minimum as we change $m_0$, again with the
constraint $P_{\rm acc}\gtrsim 0.75$.
Notice that the leftmost point corresponds to a value of $m_0-m_{\rm c}$ smaller than the ones previously
considered in this paper and therefore the comparison provides a check of the functional dependencies
on the quark mass  that we have introduced in the previous Section. 
The agreement is quite satisfactory, and always within two combined sigmas, over a large range of cost values and that gives us confidence
on the accuracy of the proposed method. More important than the discrepancy between simulations and predictions in units of combined
standard deviations is actually their relative difference, which is 10\% at most, so  the predictions are indeed quite precise.
As a final remark we point out that the dependence of the cost  on the bare subtracted quark mass turns out to be
consistent with a $1/(m_0-m_{\rm c})$ behavior. 
\section{Conclusions}
We presented a simple and general method to optimize the parameters of the 
integrators and those entering the determinant splitting in HMC simulations 
of lattice gauge theories. 
The main observation is that shadow Hamiltonians and Poisson brackets provide a clear
way to separate the dependence of (e.g.) \textcolor{black}{the violations of the energy (Hamiltonian) conservation} 
on algorithmic and integrator parameters.
The dependence on the latter is explicit, whereas that on the first is implicit in the arguments of 
the Poisson brackets. 
The idea has already been used
in the past to optimize integrators, but we have extended it here in order to simultaneously 
optimize the algorithm (i.e., tuning the Hasenbusch mass $\mu$, in the case considered), 
by showing that the dependence of the Poisson brackets on the algorithmic parameters  
is rather smooth and can be easily parameterized.
We looked at a simple case (Omelyan second order integrator, with $\alpha=1/6$, one level of Hasenbusch splitting), 
but more complicated ones can be straightforwardly
considered, at the cost of computing Poisson brackets and model their dependence
on the parameters defining the specific \textcolor{black}{factorization of the quark determinant}. 

The considered case is however relevant as a significant amount of data is available within that setup 
in the framework of non-perturbative studies of strongly interacting extensions of the Standard
Model. Those are becoming precise studies and simulations are getting computationally expensive, therefore it is crucial
to be able to re-utilize existing results in order to optimize the efficiency of the algorithms.
Since such an optimization may in principle be different for each model, a reliable and \textcolor{black}{inexpensive}
way to do it, as the one proposed here, is highly desirable.

We indeed found a very satisfactory agreement (at the ten-percent level) between 
the cost predictions from our method and actual results from simulations.

\vspace{0.25cm}
\noindent{\bf Acknowledgements.}
We thank Ari Hietanen and Martin Hansen for help and discussions in the initial phase 
of the project.
We wish to thank Tony Kennedy for useful discussions.
This work was supported
by the Danish National Research Foundation DNRF:90
grant and by a Lundbeck Foundation Fellowship grant number 2011-9799.
Local computational facilities used in this work were provided by the DeIC 
national High Performance Computing (HPC) centre at University of Southern Denmark (SDU), 
funded by SDU and the Danish e-infrastructure Cooperation (DeIC).
A.B.~acknowledges support through the Spanish MINECO project
FPA2015-68541-P, the Centro de Excelencia Severo Ochoa Programme
SEV-2016-0597 and the Ram\'on y Cajal Programme RYC-2012-10819.


\begin{thebibliography}{99}
\bibitem{Duane:1987de}
  S.~Duane, A.~D.~Kennedy, B.~J.~Pendleton and D.~Roweth,
  Phys.\ Lett.\ B {\bf 195} (1987) 216.
%
\bibitem{Hasenbusch:2001ne}
  M.~Hasenbusch,
  Phys.\ Lett.\ B {\bf 519} (2001) 177,
  [hep-lat/0107019].
%
\bibitem{Luscher:2005rx}
  M.~L\"uscher,
  Comput.\ Phys.\ Commun.\  {\bf 165} (2005) 199,
  [hep-lat/0409106].
%
\bibitem{Clark:2006fx}
  M.~A.~Clark and A.~D.~Kennedy,
  Phys.\ Rev.\ Lett.\  {\bf 98} (2007) 051601,
  [hep-lat/0608015].
%
\bibitem{Omelyanint}
I. P. Omelyan, I. M. Mryglod, and R. Folk,
Comput.\ Phys.\ Commun.\  {\bf 151} (2003) 272.
%
\bibitem{Takaishi:2005tz}
  T.~Takaishi and P.~de Forcrand,
  Phys.\ Rev.\ E {\bf 73} (2006) 036706,
  [hep-lat/0505020].
%
\bibitem{Urbach:2005ji}
  C.~Urbach, K.~Jansen, A.~Shindler and U.~Wenger,
  Comput.\ Phys.\ Commun.\  {\bf 174} (2006) 87,
  [hep-lat/0506011].
%
\bibitem{Clark:2008gh}
  M.~A.~Clark, A.~D.~Kennedy and P.~J.~Silva,
  PoS LATTICE {\bf 2008} (2008) 041
  [arXiv:0810.1315 [hep-lat]].
%
\bibitem{Clark:2011ir}
  M.~A.~Clark, B.~Joo, A.~D.~Kennedy and P.~J.~Silva,
  Phys.\ Rev.\ D {\bf 84} (2011) 071502,
  [arXiv:1108.1828 [hep-lat]].
%
\bibitem{Bussone:2016pmq}
  A.~Bussone, M.~Della Morte, V.~Drach, M.~Hansen, A.~Hietanen, J.~Rantaharju and C.~Pica,
  PoS LATTICE {\bf 2016} (2016) 260
  [arXiv:1610.02860 [hep-lat]].
%
\bibitem{Clark:2007ffa}
  M.~A.~Clark and A.~D.~Kennedy,
  Phys.\ Rev.\ D {\bf 76} (2007) 074508
  [arXiv:0705.2014 [hep-lat]].
%
\bibitem{Kennedy:2012gk}
  A.~D.~Kennedy, P.~J.~Silva and M.~A.~Clark,
  Phys.\ Rev.\ D {\bf 87} (2013) no.3,  034511
[arXiv:1210.6600 [hep-lat]].
%
\bibitem{DellaMorte:2003jj}
  M.~Della Morte {\it et al.} [ALPHA Collaboration],
  Comput.\ Phys.\ Commun.\  {\bf 156} (2003) 62,
[hep-lat/0307008].
%
\bibitem{Luscher:2010ae}
  M.~Luscher,
  arXiv:1002.4232 [hep-lat].
%
\bibitem{Creutz:1988wv}
  M.~Creutz,
  Phys.\ Rev.\ D {\bf 38} (1988) 1228.
%
\bibitem{Gupta:1990ka}
  S.~Gupta, A.~Irback, F.~Karsch and B.~Petersson,
  Phys.\ Lett.\ B {\bf 242} (1990) 437.
%
\bibitem{Clark:2010qw}
  M.~A.~Clark, B.~Joo, A.~D.~Kennedy and P.~J.~Silva,
  PoS LATTICE {\bf 2010} (2010) 323, [arXiv:1011.0230 [hep-lat]].
%
\bibitem{Meyer:2006ty}
  H.~B.~Meyer, H.~Simma, R.~Sommer, M.~Della Morte, O.~Witzel and U.~Wolff,
  Comput.\ Phys.\ Commun.\  {\bf 176} (2007) 91, [hep-lat/0606004].
%
\bibitem{DellaMorte:2008ad}
  M.~Della Morte {\it et al.} [ALPHA Collaboration],
  JHEP {\bf 0807} (2008) 037, [arXiv:0804.3383 [hep-lat]].
%
\bibitem{Hirep}
  L.~Del Debbio, A.~Patella and C.~Pica,
  Phys.\ Rev.\ D {\bf 81} (2010) 094503, 
  [arXiv:0805.2058 [hep-lat]].

\bibitem{DelDebbio:2005qa}
  L.~Del Debbio, L.~Giusti, M.~Luscher, R.~Petronzio and N.~Tantalo,
  JHEP {\bf 0602} (2006) 011, [hep-lat/0512021].

\end{thebibliography}
\end{document}